\documentclass[11pt,openany,openbib]{article}
\usepackage{amssymb}
\usepackage{amsmath}
\usepackage{epic}
\usepackage{eepic}
\usepackage{graphicx,subfigure}
\usepackage{times}
\usepackage{amsfonts}
\usepackage{eucal}
\usepackage{color}
\usepackage{pst-all}

\newcommand{\qed}{\mbox{\rule{1.6mm}{4.3mm}}}

\newtheorem{theo}{Theorem}
\newtheorem{lemma}[theo]{Lemma}
\newtheorem{co}[theo]{Corollary}

\newtheorem{defn}[theo]{Definition}
\newtheorem{proof}{Proof}
\def\proof{\textbf{Proof:~}}
\newcommand{\beqno}{ \begin{equation*} }
\newcommand{\eeqno}{ \end{equation*} }
\newcommand{\beq}{ \begin{equation} }
\newcommand{\eeq}{ \end{equation} }
\newcommand{\bR}{\boldsymbol{R}}
\newcommand{\brho}{\boldsymbol{\rho}}
\newcommand{\bx}{\boldsymbol{x}}

\usepackage{vmargin}
\setmargins{1in}{1in}{6.5in}{9in}{0pt}{0pt}{12pt}{42pt}

\begin{document}

\title{Selfish Distributed Compression over Networks: Correlation Induces Anarchy}

\author{ Aditya Ramamoorthy\footnote{Department of Electrical and Computer Engineering,
 Iowa State University,Ames, Iowa 50011. Email: adityar@iastate.edu}
\and Vwani Roychowdhury\footnote{Department of Electrical Engineering,
University of California, Los Angeles, CA 90095. \& NetSeer Inc.,
Santa Clara, CA 95054.  Email:vwani@ee.ucla.edu}
\and
Sudhir Kumar Singh\footnote{NetSeer Inc.,
Santa Clara, CA 95054. Email:suds@ee.ucla.edu}
}

\maketitle

\begin{abstract}
We consider the min-cost multicast problem (under network coding) with multiple correlated sources where each terminal
wants to losslessly reconstruct all the sources. This can be considered as the network generalization of the classical distributed source coding (Slepian-Wolf) problem.
We study the inefficiency brought forth by
the selfish behavior of the terminals in this scenario by modeling it as a noncooperative game among the terminals.
The solution concept that we adopt for this game
is the popular local Nash equilibrium (Wardrop equilibrium) adapted for the
scenario with multiple sources.
The degradation in performance due to the lack of regulation is measured
by the {\it Price of Anarchy} (POA),
which is defined as the ratio between the cost of the worst possible
Wardrop equilibrium and the
socially optimum cost. Our main result is that in contrast with the case of independent sources, the presence of source correlations can significantly increase the price of anarchy. 
Towards establishing this result we make several contributions. We characterize the socially optimal
flow and rate allocation in terms of four intuitive conditions. This result is a key technical contribution of
this paper and is of independent interest as well.
Next, we show that the Wardrop equilibrium is a socially optimal solution for a
different set of (related) cost functions. Using this, we construct explicit examples that demonstrate that the POA $> 1$ and determine near-tight upper bounds on the POA as well.
The main techniques in our analysis are Lagrangian duality theory and the usage of
the supermodularity of conditional entropy. Finally, all the techniques and results in this paper will
naturally extend to a large class of network information flow problems
where the Slepian-Wolf polytope is replaced by any contra-polymatroid (or more generally polymatroid-like set),
leading to a nice class of succinct multi-player games and allow the investigation of
other practical and meaningful scenarios beyond network coding as well.
\end{abstract}


\section{Introduction}
\label{sec:intro}

In large scale networks such as the Internet, the agents involved in producing and transmitting information often exhibit selfish behavior e.g. if a packet needs to traverse the network of various ISP's, each ISP will behave in a greedy manner and ensure that the packet spends the minimum time on its network. While this minimizes the ISP's cost it may not be the best strategy from a overall network cost perspective. Selfish routing, that deals with the question of network performance under a lack of regulation has been studied extensively (see \cite{AGTbook, roughgarden05}) and has developed as an area of intense research activity. However, by and large most of these studies have considered the network traffic injected into the network at various sources to be independent.

From an information theoretic perspective there is no need to consider the sources involved in the transmission to be independent. In this work we initiate the study of network optimization issues related to the transmission of correlated sources over a network when the agents involved are selfish. In particular, we concentrate on the problem of multicasting correlated sources over a network to different terminals, where each terminal is interested in losslessly reconstructing all the sources. We assume that the network is capable of network coding. Under this scenario, a generalization of the classical Slepian-Wolf theorem of distributed source coding \cite{tracey_ciss} holds for arbitrary networks. In particular, when the network performs random linear network coding each terminal can recover the sources under appropriate conditions on the Slepian-Wolf region and the capacity region of the terminals with respect to the sources, thereby allowing distributed source coding over networks. 
The selfish agents in our set-up are the terminals who pay for the resources. Each terminal aims to minimize her own cost while ensuring that she can satisfy her demands. It is important to note that this is a generalization of the problem of minimum cost selfish multicast of independent sources considered by Bhadra et al. \cite{BhadraSG06}.

\subsection{Our Results}
In this work, we model the scenario as a noncooperative game amongst the selfish terminals who
request rates from sources and flows over network paths such that their individual cost is minimized (i.e. with no regard for social welfare) while allowing for reconstruction of all the sources. We investigate properties of the socially optimal solution and define appropriate solution concepts (Nash equilibrium and Wardrop equilibrium) for this game and investigate properties of the flow-rates at equilibrium. We briefly describe our contributions below.\\
\begin{itemize}
\item[i)] \textit{Characterization of social-optimality conditions.} The problem of computing the socially optimal cost is a convex program. We present a precise characterization of the optimality conditions of this convex program in terms of four intuitive conditions, using Lagrangian duality theory and by judiciously exploiting the super-modularity of conditional entropy. 
This result is a key technical contribution of
this paper and is of independent interest as well. \\
\item[ii)] \textit{Demonstrating the equivalence of flow-rates at equilibrium with social-optimal solutions for alternative instances.} We consider certain meaningful market models that split resource costs amongst the different terminals and show that the flows and rates under the game-theoretic equilibriums are in fact socially optimal solutions for a different set of cost functions. This characterization allows us to quantify the degradation caused by the lack of regulation. The measure of performance degradation due to such loss in regulation that we adopt is the \textit{Price of Anarchy} (POA), which is defined as the ratio between the cost of the worst possible equilibrium and the socially optimum cost \cite{KP99,papa01,RT02,roughgarden05}.\\
\item[iii)]
 \textit{ Showing that source correlation induces anarchy.} The main result of this work is that the presence of source correlations can significantly increase the POA under reasonable cost-splitting mechanisms. This is in stark contrast to the case of multicast with independent sources, where for a large class of cost functions, cost-splitting mechanisms can be designed that ensure that the price of anarchy is one. We construct explicit examples where the POA is greater than one and also obtain an upper bound on the POA which is near tight.
\end{itemize}

Finally, we expect that the techniques developed in the present work will be applicable to a large class of network information flow problems with correlated sources where the Slepian-Wolf polytope is replaced by polymatroid-like objects. These include multi-terminal source coding with high resolution \cite{zamirB99} and the CEO problem \cite{prabhatse}.

\subsection{Background and Related Work}
Distributed source coding (or distributed compression) (see \cite{coverthomas}, Ch. 14 for an overview) considers the problem of compressing multiple discrete memoryless sources that are observing correlated random variables.
The landmark result of Slepian and Wolf \cite{slepianwolf} characterizes the feasible rate region for the recovery of the sources.
However, the problem of Slepian and Wolf considers a direct link between the sources and the terminal. More generally one would expect that the sources communicate with the terminal over a network. Different aspects of the Slepian-Wolf problem over networks have been considered in (\cite{barrosservetto,cristescuBV05,ramamoorthy07}). Network coding (first introduced in the seminal work of Ahlswede et al. \cite{al}) for correlated sources was studied by Ho et al. \cite{tracey_ciss}. They considered a network with a set of sources and a set of terminals and showed that as long as the minimum cuts between all non-empty subsets of sources and a particular terminal were sufficiently large (essentially as long as the Slepian-Wolf region of the sources has an intersection with the capacity region of a given terminal), random linear network coding over the network followed by appropriate decoding at the terminals achieves the Slepian-Wolf bounds.

The problem of minimum cost multicast under network coding has been addressed in the work of \cite{lunmincost,LiL05}. The multicast problem has also been examined by considering selfish agents \cite{BhadraSG06,Li07,Li08}. Our work is closest in spirit to the analysis of Bhadra et al. \cite{BhadraSG06} that considers selfish terminals.
In this scenario, for a large class of edge cost functions,
they develop a pricing mechanism for allocating the edge costs among the different terminals
and show that it leads to a globally optimal solution to the original optimization problem, i.e. the price of anarchy is one.
Their POA analysis is similar to that in the case of selfish routing \cite{RT02,roughgarden05}.
Our model is more general and our results do not generalize from theirs in a straightforward manner.
In particular, we need to judiciously exploit several non-trivial properties of the Slepian-Wolf polytope
in our analysis.

Further, motivated by the need to deal with selfish users, particularly in network setting, there has been a large
body of recent work at the intersection of networking, game theory, economics, and theoretical computer science \cite{AGTbook,BV06,HV07}. This work adds another interesting dimension to this interdisciplinary area.

\section{The Model}
\label{model}

Consider a directed graph $G = (S \cup T \cup V, E)$.
There is a set of source nodes $S$ that may be
correlated and a set of sinks $T$ that are the
terminals (i.e. receivers). Each source node observes a discrete memoryless source
$X_i$. The Slepian-Wolf region of the sources is assumed to be
known and is denoted $\mathcal{R_{SW}}$. For notational simplicity,
let $N_S=|S|, N_T=|T|$,
$S=\{1,2,\dots,N_S\}$, and $T=\{t_1,t_2,\dots,t_{N_T}\}$.
The set of paths from source $s$ to terminal $t$ is denoted by
$\mathcal{P}_{s,t}$. Further, define $\mathcal{P}_t = \cup_{s \in S} \mathcal{P}_{s,t}$ i.e.
the set of all possible paths going to terminal $t$, and $\mathcal{P} = \cup_{t \in T} \mathcal{P}_t$,
the set of all possible paths.
A \textit{ flow} is an assignment of non-negative reals to each path $P
\in \mathcal{P}$.
The flow on $P$ is denoted $f_P$.
A \textit{ rate} is a function $ R: S \times T \longrightarrow \mathcal{R}^{+}$, i.e.
the rate requested by the terminal $t$ from the source $s$ is $R_{s,t}$.
We will refer to a flow and rate pair $(f,R)$ as \emph{flow-rate}.
Also, let us denote the rate vector for terminal $t$ by $\bR_t$ and the
vector of requested rates at source $s$ by $\brho_s$ i.e. $\bR_t=(R_{1,t}, R_{2,t}, \dots, R_{N_S,t})$
and $\brho_s=(R_{s,t_1}, R_{s,t_2}, \dots, R_{s,t_{N_T}})$.

Associated with each edge $e \in E$ is a cost $c_e$, which takes as argument
a scalar variable $z_e$ that depends on the flows to various terminals passing through $e$.
Similarly, let $d_s$ be the cost function corresponding to the source $s$, which takes as argument
a scalar variable $y_s$ that depends on the rates that various terminals request from $s$.
These functions $c_e$'s and $d_s$'s are assumed to be \textit{ convex, positive, differentiable
and monotonically increasing}. Further, the functions $\int \frac{c_e(x)}{x} ~ dx$ are
also convex, positive, differentiable
and monotonically increasing. In particular, these conditions are
satisfied by functions like $x^a, a > 1$ and $x
e^{bx}, b > 0$ among others.

The network connection we are interested in supporting is one where each terminal can reconstruct all the sources.
i.e. we need to jointly allocate rates and flows for each terminal so
that it can reconstruct the sources. We now present a formal
description of the optimization problem under consideration.

\subsection{Min-Cost Multicast with Multiple Sources}
Let us call the quadruple $(G,c,d,\mathcal{R_{SW}})$ an \emph{instance}. The problem of minimizing the total cost for the instance $(G,c,d,\mathcal{R_{SW}})$ can be formulated as
\begin{eqnarray}
& \text{minimize~} & C(f,R)  = \sum_{e \in E} c_e(z_e) + \sum_{s \in S} d_s(y_s) \nonumber \\
&\text{subject to~}
 & f_P  \geq 0 ~~ \forall P \in \mathcal{P} \nonumber \\
 \label{cons_3}& (NIF-CP) ~~  & \sum_{P \in \mathcal{P}_{s,t}} f_P  \geq R_{s,t}
~~  \forall s \in S , \forall t \in T\\
& & \bR_t  \in \mathcal{R_{SW}}~~ \forall t \in T \nonumber
\end{eqnarray}
where $z_e, \forall e \in E$ is a function of $x_{e,t_1}, x_{e,t_2},
\dots, x_{e,t_{N_T}}$, that we denote
$z_e(x_{e,t_1}, x_{e,t_2}, \dots, x_{e,t_{N_T}})$ with $x_{e,t}  = \sum_{P \in \mathcal{P}_t: e
  \in P} f_P ~~ \forall e \in E , ~~ \forall t \in T$,
and $y_s, \forall s \in S$ is a function of $\brho_s$ that we will denote $y_s(\brho_s)$.

The formulation above is similar to the one presented in \cite{BhadraSG06}. However since we consider source correlations as well, their formulation is a specific case of our formulation.
Since network coding allows the sharing of
edges, the penalty at an edge is only the maximum and not the sum i.e.
$z_e$ is the maximum flow (among the different terminals)
across the edge $e$. Similarly,
the penalty at the sources for higher resolution quantization is
also driven by the maximum level requested by each terminal i.e.
$y_s$ is also maximum.
In this work, for differentiability requirements
the maximum function will be approximated as $L_p$ norm with a large $p$. Nevertheless, most of our analysis is done where
$z_e$ and $y_s$ are non-decreasing functions partially differentiable
with respect to their arguments,
such that $c_e(z_e)$ and $d_s(y_s)$ are convex, positive, differentiable and monotonically increasing. Note that in the formulation above, the objective function is convex and all constraints are linear which implies that this is a convex optimization problem.

The constraint
(\ref{cons_3}) above models the fact that the total flow from the
source $s$ to a terminal $t$ needs to be at least
$R_{s,t}$. Finally, the rate point of each terminal
$\bR_t$ needs to be within the Slepian-Wolf
polytope. A flow-rate $(f,R)$ satisfying all the
conditions in the above optimization problem (i.e. \emph{(NIF-CP)} ) will be called a \textit{ feasible} flow-rate for
the instance $(G,c,d,\mathcal{R_{SW}})$ and the cost $C(f,R)$ will be
referred to as the \emph{social cost} corresponding to this flow-rate.
Also, we will call a solution $(f^{*},R^{*})$ of the above
problem as an \emph{OPT} flow-rate for the instance $(G,c,d,\mathcal{R_{SW}})$.

Consider a feasible flow-rate $(f,R)$ for the above optimization
problem. It can be seen that the value of the flow from $A \subseteq
S$ to a terminal $t \in T$ is $\sum_{P \in \cup_{s \in A}
  \mathcal{P}_{s,t}} f_P \geq \sum_{s \in A} R_{s,t}$. Since
$\mathbf{R}_t \in \mathcal{R_{SW}}$ the result of \cite{tracey_ciss}
shows that random linear network coding followed by appropriate
decoding at the terminals can recover the sources with high
probability. Conversely the result of \cite{han80,barrosservetto} shows the necessity of the
existence of such a flow.

\subsection{Terminals' Incentives and the Distributed Compression Game}

The above formulation for social cost minimization for the instance $(G,c,d,\mathcal{R_{SW}})$ disregards the fact that the agents who pay for the costs
incurred at the edges and the sources may not be cooperative and may have incentives for strategic manipulation.
In this work we consider the scenario where the terminals pay for the network resources
they are being provided. The terminals are noncooperative and will behave selfishly trying
to minimize their own respective costs without regard to the social cost, while
ensuring that they can reconstruct all the sources. We have the
following assumptions.

\begin{itemize}

\item[(i)] Let $(f, R)$ denote a feasible flow rate for the instance
$(G,c,d,\mathcal{R_{SW}})$. The network operates via random linear
network coding (or some practical linear network coding scheme) over the subgraph of $G$ induced by the corresponding
$\{z_e\}$ for $e \in E$. The terminals are capable of performing
appropriate decoding to recover the sources.

\item[(ii)] Each terminal $t \in T$ can request for any specific set of flows on the paths $P \in \mathcal{P}_t$ and rates $\mathbf{R}_t$ as long as such a request allows reconstruction of the sources at $t$. There is a mechanism in the network by means of which this request is accommodated i.e. the subgraph over which random linear network coding is performed is adjusted appropriately.

\end{itemize}

In this work we wish to characterize flow-rates that represent an equilibrium
among selfish terminals who act strategically to minimize their own costs. Furthermore, we shall systematically study the loss that occurs due to the mismatch between the social goals and terminal's selfish goals.

Towards this end, we now formally model the game originating from the selfish
behavior of the terminals. We model this game as a \textit{normal formal game} or \emph{strategic game} \cite{ORgamebook} ,
which we refer to as the \textit{Distributed Compression Game(DCG)}.

A normal form game, denoted $(\mathcal{N}, \{A_i\}_{i \in \mathcal{N}}, \{\succeq_i\}_{i \in \mathcal{N}} )$,
consists of the set of \textit{players} $\mathcal{N}$, the tuple of \textit{set of strategies} $A_i$
for each player $i \in \mathcal{N}$,
and the tuple of \textit{preference relations} $\succeq_i$ for each player $i \in \mathcal{N}$
on the set $\mathcal{A} = \times_{i \in \mathcal{N}} A_i $. For $a,b \in \mathcal{A}$, $ a \succeq_i b $ means that the player $i$ prefers
the tuple of strategies $a$ to the tuple of strategies $b$. In the context of \textit{Distributed Compression Game},
given an instance $(G,c,d,\mathcal{R_{SW}})$, these parameters are defined as follows.

\subsubsection{The Distributed Compression Game}

\begin{itemize}
\item \textbf{Players:} $\mathcal{N}= T$, i.e. the terminals are the players. This is because, as mentioned above,
the terminals are the users and they are the ones who pay for the network resources they are being provided.

\item \textbf{Strategies:} The strategy set of a player $t \in T$ consists of tuples $(f_t,\bR_t)$ where

\begin{itemize}
\item $f_t$ is the vector of flows on paths going to $t$, i.e. the vector of values $f_P$ for all $P \in \mathcal{P}_t$,
and recall that $\bR_t$ denotes the rate vector for terminal $t$;

\item  $f_P  \geq 0 ~~ \forall P \in \mathcal{P}_t , \sum_{P \in \mathcal{P}_{s,t}} f_P  \geq R_{s,t} ~~  \forall s \in S$ and
 $\bR_t  \in \mathcal{R_{SW}}$.
\end{itemize}

Therefore,
\begin{equation}
A_t = \left\{  (f_t,\bR_t): \begin{array}{l} f_P  \geq 0 ~~ \forall P \in \mathcal{P}_t, \\
 \sum_{P \in \mathcal{P}_{s,t}} f_P  \geq R_{s,t} ~~  \forall s \in S, \\
 \bR_t  \in \mathcal{R_{SW}}
\end{array}
\right\}.
\end{equation}
Note that a feasible flow-rate $(f,R)$ for the instance $(G,c,d,\mathcal{R_{SW}})$ is
an element of the set $\mathcal{A}=\times_{t \in T} A_t$ defined for the same instance.

\item \textbf{Preference Relations:}
 To specify the preference relation of terminal $t \in T$, we need to know how much
does she pay given a feasible flow-rate $(f,R)$ i.e. what fractions of the costs at
various edges and sources are being paid by $t$? To this end,
we need market models, i.e. mechanisms for splitting the
costs among various terminals.

\begin{itemize}
\item \textit{ Edge Costs:} At a flow $f$, the cost of an edge  $e \in E$ is $c_e(z_e)$.
It is split among the terminals $t \in T$, each paying a fraction of this cost.
Let us say that the fraction paid by the player $t$ is $\Psi_{e,t}(\bx_e)$ i.e. the player $t$ pays
$c_e(z_e) \Psi_{e,t}(\bx_e)$ for the edge $e$ where $\bx_e$ denotes the vector
$(x_{e,t_1}, x_{e,t_2}, \dots, x_{e,t_{N_T}})$.
Of course, $\sum_{t \in T}\Psi_{e,t}(\bx_e) = 1$ to ensure that the
total cost is borne by someone or the other. The total cost borne by $t$ across all the edges
is $\sum_{e \in E} c_e(z_e) \Psi_{e,t}(\bx_e)$, denoted
$C_{E}^{(t)}(f)$.

\item \textit{ Source Costs:} At a rate $R$, the cost for the source $s$ is $d_s(y_s)$, which is split
 among the terminals $t \in T$, such that $t$ pays
a fraction $\Phi_{s,t}(\brho_s)$ i.e. the player $t$ pays
$d_s(y_s) \Phi_{s,t}(\brho_s)$ for the source $s$. Of course, $\sum_{t \in T}\Phi_{s,t}(\brho_s) = 1$.
Therefore, the total cost borne by $t$ for all sources, denoted $C_S^{(t)}(R)$, is
$\sum_{s \in S} d_s(y_s) \Phi_{s,t}(\brho_s)$.
\end{itemize}

Thus, with the \textit{edge-cost-splitting mechanism} $\Psi$ and the \textit{source-cost-splitting mechanism} $\Phi$,
the total cost incurred by the player $t \in T$ at flow-rate $(f,R)$ denoted $C^{(t)}(f,R)$ is
\begin{eqnarray*}
C^{(t)}(f,R) &=&C_{E}^{(t)}(f) + C_S^{(t)}(R) \nonumber \\
 &=& \sum_{e \in E} c_e(z_e) \Psi_{e,t}(\bx_e) + \sum_{s \in S} d_s(y_s) \Phi_{s,t}(\brho_s).
\end{eqnarray*}

Now, each terminal $t$ would like to minimize its own cost i.e. the function $C^{(t)}(f,R)$ and therefore the preference
relations $\{\succeq_t\}$ are as follows. For two flow-rates $(f,R) \in \mathcal{A}$ and $(\tilde{f},\tilde{R}) \in \mathcal{A}$,
$(f,R) \succeq_t (\tilde{f},\tilde{R})$ if and only if $C^{(t)}(f,R) \leq C^{(t)}(\tilde{f},\tilde{R})$. Also,
$(f,R) \succ_t (\tilde{f},\tilde{R})$ iff $C^{(t)}(f,R) < C^{(t)}(\tilde{f},\tilde{R})$.
\end{itemize}

Note that for specifying a Distributed Compression Game, in addition to the parameters $G,c,d$ and $\mathcal{R_{SW}}$ we
also need the cost-splitting mechanisms $\Psi$  and $\Phi$.
We will call $(G,c,d,\mathcal{R_{SW}},\Psi,\Phi)$
as an instance of the Distributed Compression Game.

\subsubsection{Solution Concepts for the Distributed Compression Game}
We now outline the possible solution concepts in our scenario. These are essentially dictated by the level of sophistication of the terminals. Sophistication refers to the amount of information and computational resources available to a terminal. In this work we shall work with two different solution concepts that we now discuss.

a) \textit{ Nash Equilibrium.} The solution concept of Nash equlibrium requires the complete information setting and requires each terminal
to compute her best response to any given tuple of strategies of the other players.
For notational simplicity, let
$f_{-t}$ be the vector of flows on paths not going to terminal $t$
i.e. the vector of values $f_P$ for all $P \in \mathcal{P}-\mathcal{P}_t$, therefore $f = (f_{-t},f_t)$. Similarly,
$\bR_{-t}$ is the vector of rates corresponding to all players other than $t$, therefore $R = (\bR_{-t},\bR_t)$.
In our setting, the best response problem of a terminal $t$ is to minimize her cost function
$C^{(t)}(f_{-t},f_t, \bR_{-t},\bR_t)$ over $(f_t,\bR_t) \in A_t$ given any $(f_{-t},\bR_{-t})$. Therefore a Nash flow-rate is defined as follows.
\begin{defn}
\emph{\textbf{ (Nash flow-rate)}} A flow-rate $(f,R)$ feasible for the instance
$(G,c,d,\mathcal{R_{SW}})$ is at Nash equilibrium,
or is a Nash flow-rate for instance
$(G,c,d,\mathcal{R_{SW}},\Psi,\Phi)$, if
 $\forall t \in T$,
\begin{equation*}
C^{(t)}(f,R)  \leq C^{(t)}(f_{-t},\tilde{f}_t, \bR_{-t},\tilde{\bR}_t)  ~~ \forall  (\tilde{f}_t, \tilde{\bR}_t) \in A_t.
\end{equation*}
\end{defn}
We note that computing the best response will in general require a
given terminal to know flow assignments on all possible paths and rate
vectors for all the terminals. Moreover, convexity of the objective
function in $NIF-CP$ (i.e. social cost $C(f,R)$) does not imply convexity of $C^{(t)}(f_{-t},f_t,
\bR_{-t},\bR_t)$ in the variables $(f_t,\bR_t) \in A_t$ in general. Therefore the
computational requirements at the terminals may be large.
Consequently Nash equilibrium does not seem to be an
appropriate solution concept for the Distributed Compression Game when viewed through the algorithmic
lens.

b) \textit{ Wardrop Equilibrium.}
From a practical standpoint, a terminal
may only have partial knowledge of the system and may be
computationally constrained.
A solution concept more appropriate under such situations is that of local Nash equilibrium or Wardrop equilibrium that is widely adopted in selfish routing and transportation literature \cite{roughgarden05,BMW56,DS69}. We note that this solution concept has also been utilized in \cite{BhadraSG06} and is further justified in \cite{Gupta97asystem}.
We first present the precise definition of the Wardrop equilibrium in our case and then provide an intuitive justification.
Towards this end, we need to define the marginal cost of a path.
\begin{defn}
\emph{\textbf{ (Marginal Cost of a Path)}}
For a  $P \in \mathcal{P}_t$ its marginal cost is
\begin{equation*}
C_P(f) = \sum_{e \in P}  \frac{c_e(z_e) \Psi_{e,t}(\bx_e)}{x_{e,t}}.
\end{equation*}
\end{defn}
Therefore, for the terminal $t$, the total cost for the edges,
$C_E^{(t)}$, can be equivalently written as
\begin{equation*}
C_E^{(t)}(f) = \sum_{P \in \mathcal{P}_t} C_P(f) f_P.
\end{equation*}

\begin{defn}
\label{waldropdef}
\emph{\textbf{ (Wardrop flow-rate)}} A flow-rate $(f,R)$ feasible for the instance
$(G,c,d,\mathcal{R_{SW}})$ is at local Nash equilibrium,
or is a Wardrop flow-rate for instance
$(G,c,d,\mathcal{R_{SW}},\Psi,\Phi)$, if it satisfies the following
conditions.
\begin{enumerate}

\item $\forall t \in T, ~ ~ \forall s \in S$, we have
\begin{equation*}
\sum_{P \in \mathcal{P}_{s,t}} f_P = R_{s,t}.
\end{equation*}

\item $\forall t \in T$, we have
\begin{equation*}
\sum_{s \in S} R_{s,t} = H(X_S).
\end{equation*}

\item  $\forall t \in T, ~ \forall s \in S$,
$P, Q \in \mathcal{P}_{s,t}$ with $f_{P} > 0$,
\begin{equation*}
C_{P}(f) \leq C_{Q}(f).
\end{equation*}

\item For $t \in T $, let $j \in S$ participates in \textbf{all tight} rate inequalities involving
$i \in S$ (i.e. if $A \subseteq S$, such that $i \in A$ and $\sum_{l \in A} R_{l,t} = H(X_A | X_{-A})$\footnote{We use $H(X_A | X_{-A})$ and $H(X_A | X_{A^c})$ interchangeably in the text to denote the joint entropy of the sources in set $A$ given the remaining sources.}, then $j \in A$) and
let $P \in \mathcal{P}_{i,t}, Q \in \mathcal{P}_{j,t}$ with $f_P > 0$ then we have
\begin{equation*}
C_{P}(f) + \frac{\partial C_S^{(t)}(R)}{\partial R_{i,t}}  \leq C_{Q}(f) + \frac{\partial C_S^{(t)}(R)}{\partial R_{j,t}}.\\
\end{equation*}
\end{enumerate}
\end{defn}

Intuitively, conditions (1) and (2) require that each terminal
requests as little rate and flow as possible. Condition (3) ensures
that an infitesimally small change in flow allocations from path $P$
(where $f_P > 0$) to path $Q$ where $P, Q \in \mathcal{P}_{s,t}$, will
increase the sum cost along paths in $\mathcal{P}_t$. Now, consider an
infitesimally small change in flow allocation from $P \in
\mathcal{P}_{i,t}$ (where $f_P > 0$) to $Q \in
\mathcal{P}_{j,t}$. This also requires a corresponding change in the
rates requested from sources $i$ and $j$ by terminal $t$. Under
certain constraints on the source $j$, Condition (4) ensures that the
overall effect of this change will serve to increase terminal $t$'s
cost.
The conditions on the source $j$ are well-motivated in light of the
characterization of Nash flow-rate in
section \ref{nash-flow-rate-prop-sec}
in the case when the best response problem of every terminal is convex.

We remark that a Nash flow-rate may not always be a Wardrop flow-rate and vice versa.
When sources are independent, condition (2) implies that $R_{s,t}=H(X_s)$ for
all $s \in S, t \in T$ and it is not required to check the condition (4). Also we can recover condition (3) by setting $i = j$ in condition (4). They are stated separately for the sake of clarity.

As we discussed earlier, the solution concept based on Wardrop equilibrium seems more suitable
to our scenario and consequently we define the price of anarchy \cite{KP99,papa01,roughgarden05} in terms of Wardrop flow-rate instead
of Nash flow-rate.

\begin{defn}
\label{poadef}
\emph{\textbf{ Price of Anarchy(POA)}:}
 Let  $\mathcal{C}$ be a class of edge cost functions,
$\mathcal{D}$ be a class of source cost functions, $\mathcal{G}$ be a class of networks/graphs,
$\Psi$ be an edge cost splitting mechanism,  $\Phi$ be a source cost splitting mechanism,
and $\mathcal{M}$ be a set of Slepian-Wolf polytopes. We will refer to $(\mathcal{G}, \mathcal{C},\mathcal{D},\Psi,\Phi,\mathcal{M})$ as a \emph{scenario}.
The price of anarchy for the scenario $(\mathcal{G}, \mathcal{C},\mathcal{D},\Psi,\Phi,\mathcal{M})$, denoted
$\rho(\mathcal{G},\mathcal{C},\mathcal{D},\Psi,\Phi,\mathcal{M})$,
is defined as maximum over all instances $(G,c,d,\mathcal{R_{SW}})$ with $G \in \mathcal{G}, c \in \mathcal{C}, d \in \mathcal{D}, \mathcal{R_{SW}} \in \mathcal{M}$,
of the ratio between the cost of worst possible Wardrop flow-rate for the instance
$(G,c,d,\mathcal{R_{SW}},\Psi,\Phi)$ and the cost of OPT flow-rate (i.e. the socially optimal cost)
for the instance $(G,c,d,\mathcal{R_{SW}})$. That is,

\begin{align*}
 \rho(\mathcal{G}, \mathcal{C},\mathcal{D},\Psi,\Phi,\mathcal{M})
=  \max_{G \in \mathcal{G}, c \in \mathcal{C}, d \in \mathcal{D}, \mathcal{R_{SW}} \in \mathcal{M}} \left(\frac{\max_{\textrm{$(f,R)$ is a Wardrop flow-rate for $(G,c,d,\mathcal{R_{SW}},\Psi,\Phi)$}} ~ C(f,R) }{C_{OPT}(G,c,d,\mathcal{R_{SW}})}
\right),
\end{align*}

where $C_{OPT}(G,c,d,\mathcal{R_{SW}})$ refers to the optimal cost of $NIF-CP$ for the instance $(G,c,d,\mathcal{R_{SW}})$.
\end{defn}

Let us denote the set of Slepian-Wolf polytopes corresponding to the case where there are no source correlations (i.e.
$H(X_A|X_{-A})=H(X_A)$ for all $A \in S$) by $\mathcal{M}_{ind}$ (subscript $ind$ denotes - \textit{ independent}) and the set of  Slepian-Wolf polytopes
corresponding to the case where sources are correlated (i.e. there exists $A \subseteq S$ with
$H(X_A|X_{-A}) < H(X_A)$) by $\mathcal{M}_c$.
Also, we use $\mathcal{G}_{all}$ to denote the class of all graphs where every $t \in T$ is connected to every $s \in S$,
and $\mathcal{G}_{dsw}$ (subscript $dsw$ denotes - \textit{ direct Slepian-Wolf}) to denote the class of complete bipartite graphs between the set of sources and the set of terminals.
Note that $\mathcal{G}_{dsw}$ corresponds to the case where every terminals is directly connected to every source by an edge
and no network coding is required.
A question we will be most concerned
with in this work is whether
$\rho(\mathcal{G}, \mathcal{C},\mathcal{D},\Psi,\Phi,\mathcal{M}_c) > \rho(\mathcal{G}, \mathcal{C},\mathcal{D},\Psi,\Phi,\mathcal{M}_{ind})$,
and in particular whether $\rho(\mathcal{G}, \mathcal{C},\mathcal{D},\Psi,\Phi,\mathcal{M}_c) > 1$
but $\rho(\mathcal{G}, \mathcal{C},\mathcal{D},\Psi,\Phi,\mathcal{M}_{ind})=1$ for meaningful classes of cost functions $\mathcal{C}, \mathcal{D}$ and reasonable splitting mechanisms $\Psi$ and $\Phi$ i.e. does correlation induce anarchy?

\section{Some Properties of Slepian-Wolf Polytope }
\label{app-sw-prop}
In this section, we establish two properties of Slepian-Wolf polytope that will be useful
in the latter sections.

\begin{lemma}
\label{int-uni-rate-tight}
Let $\bR_t \in \mathcal{R_{SW}}$ i.e. $\sum_{l \in A} R_{l,t} \geq H(X_A|X_{-A})$ for all $A \subseteq S$.
If $S_1,S_2 \subseteq S$ satisfy
$$\sum_{l \in S_1} R_{l,t} = H(X_{S_1}|X_{-{S_1}})$$ and
$$\sum_{l \in S_2} R_{l,t} = H(X_{S_2}|X_{-{S_2}})$$ then we have
$$\sum_{l \in  S_1 \cap S_2 } R_{l,t} = H(X_{ S_1 \cap S_2}|X_{- (S_1 \cap S_2)})$$
and
$$\sum_{l \in  S_1 \cup S_2 } R_{l,t} = H(X_{ S_1 \cup S_2}|X_{- (S_1 \cup S_2)}).$$
\end{lemma}
\proof We have,
\begin{align*}
\sum_{l \in  S_1 \cap S_2 } R_{l,t} + \sum_{l \in  S_1 \cup S_2 }
R_{l,t}  & = \sum_{l \in S_1} R_{l,t} + \sum_{l \in S_2} R_{l,t} \\
 & =  H(X_{S_1}|X_{-{S_1}}) +  H(X_{S_2}|X_{-{S_2}}) \\
& \leq  H(X_{ S_1 \cap S_2}|X_{- (S_1 \cap S_2)}) + H(X_{ S_1 \cup S_2}|X_{- (S_1 \cup S_2)})
\end{align*}
where in the second step we have used the supermodularity property
of conditional entropy.
Now we are also given that $$\sum_{l \in  S_1 \cap S_2 } R_{l,t} \geq  H(X_{ S_1 \cap S_2}|X_{- (S_1 \cap S_2)})$$ and
 $$\sum_{l \in  S_1 \cup S_2 } R_{l,t} \geq H(X_{ S_1 \cup S_2}|X_{-
   (S_1 \cup S_2)}).$$
Therefore we can conclude that
$$\sum_{l \in  S_1 \cup S_2 } R_{l,t} = H(X_{ S_1 \cup S_2}|X_{- (S_1 \cup S_2)})$$ and
$$\sum_{l \in  S_1 \cap S_2 } R_{l,t} = H(X_{ S_1 \cap S_2}|X_{- (S_1 \cap S_2)}).$$
\endproof

\begin{theo}
\label{thm:sum-rate-constraint} Consider a vector $(R_1, R_2,
\dots, R_n)$ such that \beqno
\begin{split} \sum_{i \in A} R_i &\geq H(X_A |X_{A^c}), \text{~for all $A
\subset \{1, 2, \dots , n\}$, and}\\ \sum_{i = 1}^n R_i &> H(X_1,
X_2, \dots, X_n). \end{split}\eeqno Then there exists another
vector $(R_1^{'}, R_2^{'}, \dots, R_n^{'})$ such that $R_i^{'}
\leq R_i$ for all $i = 1, 2, \dots n$ and \beqno
\begin{split} \sum_{i \in A} R_i^{'} &\geq H(X_A |X_{A^c}), \text{~for all $A
\subset \{1, 2, \dots , n\}$, and}\\ \sum_{i = 1}^n R_i^{'} &=
H(X_1, X_2, \dots, X_n). \end{split}\eeqno
\end{theo}
\emph{Proof}. We claim that there exists a $R_{j^*} \in \{R_1,
R_2, \dots, R_n\}$ such that all inequalities in which $R_{j^*}$
participates are loose. 
The proof of this claim follows.
\par Suppose that the above claim is not true. Then for all $R_i$ where $i \in
\{ 1, 2, \dots, n\}$, there exists at least one subset $S_i
\subset \{1, 2, \dots, n\}$ such that, \beqno \sum_{k \in S_i} R_k
= H(X_{S_i} | X_{S_i^c}). \eeqno i.e. each $R_i$ participates in at
least one inequality that is tight.

Now by applying Lemma \ref{int-uni-rate-tight} on the sets $S_1,S_2,\dots,S_n$, since
 $S_1 \cup S_2 \dots \cup S_n = \{1, 2, \dots, n\}$, we get
$\sum_{i = 1}^n R_i = \sum_{i \in S_1 \cup S_2 \dots \cup S_n} R_i= H(X_{S_1 \cup S_2 \dots \cup S_n}|X_{-(S_1 \cup S_2 \dots \cup S_n)})= H(X_1, X_2, \dots, X_n)$,
which is a contradiction.

The above argument shows that there exists some $j^*$ such
that all inequalities in which $R_{j^*}$ participates are loose.
Therefore we can reduce $R_{j^*}$ to a new value $R^{red}_{j^*}$
until one of the inequalities in which it participates is tight.
If the sum-rate constraint is met with equality then we can set
$R_{j^*}^{'} = R^{red}_{j^*}$ otherwise we can recursively apply
the above procedure to arrive at a new vector that is component-wise
smaller that the original vector $(R_1, R_2, \dots, R_n)$.
\endproof

\section{Characterizing the Optimal Flows and Rates}
\label{opt-flow-rate-problem}
In this section, we investigate the properties of an \emph{OPT} flow-rate
via Lagrangian duality theory \cite{boyd_van}.
Since the optimization problem \emph{(NIF-CP)} is convex and the constraints
are such that the strong duality holds, the \textit{ Karush-Kuhn-Tucker(KKT)} conditions exactly characterize optimality \cite{boyd_van}.
Therefore, we start out by writing the Lagrangian dual of \emph{NIF-CP},
\begin{align*}
L &= \sum_{e \in E} c_e(z_e) + \sum_{s \in S} d_s(y_s) -
\sum_{P \in \mathcal{P}} \mu_{P} f_P 
+ \sum_{s \in S}  \sum_{t \in
T} \lambda_{s,t} (R_{s,t} - \sum_{P \in \mathcal{P}_{s,t}} f_P)  \\
& ~~ + \sum_{t \in T} \left[ \sum_{A \subseteq S} \nu_{A,t} 
\left(H(X_A |X_{A^c}) - \sum_{i \in A} R_{i,t}\right)\right]
\end{align*}
where $\mu_{P} \geq 0, \lambda_{s,t} \geq 0$ and $\nu_{A,t} \geq 0$ are the
dual variables (i.e. Lagrange multipliers).
For notational simplicity, let us denote  the
partial derivative of $z_e$ with respect to $x_{e,t}$, $\frac{\partial z_e}{\partial x_{e,t}}$ by $z_{e,t}^{'}$.
Note that the partial derivative of $x_{e,t}$ w.r.t. to
$f_{P}$ is $1$ for a $P \in \mathcal{P}_t$. Similarly, we denote
the partial derivative of $ y_s$ with respect to $ R_{s,t}$, $\frac{\partial y_s}{\partial R_{s,t}}$  by $y_{s,t}^{'}$.
The \emph{KKT} conditions are then given by the following equations that hold $\forall ~ s \in S, t \in T$,
\begin{align}
\frac{\partial L}{\partial f_{P}} &= \sum_{e \in P}
c_e^{'}(z_e) z_{e,t}^{'}(\bx_e) - \mu_{P} - \lambda_{s,t} = 0,
~ \forall P \in \mathcal{P}_{s,t}, \text{~and}  \label{kkt-flow}
\end{align}
\begin{align}
\frac{\partial L}{\partial R_{s,t}} &= d_s^{'}(y_s)
y_{s,t}^{'}(\brho_s) + \lambda_{s,t} - \sum_{A \subseteq
S:s \in A} \nu_{A,t} = 0  \label{kkt-flow-rate}
\end{align}
along with the feasibility of
the flow-rate $(f,R)$ and the complementary slackness conditions, $\mu_{P} f_P = 0$
for all $P \in \mathcal{P}$, $\lambda_{s,t} (R_{s,t} - \sum_{P \in \mathcal{P}_{s,t}} f_P) = 0$
for all $ s \in S,
t \in T$, and $\nu_{A,t}
\left(H(X_A |X_{A^c}) - \sum_{i \in A} R_{i,t}\right)  = 0$ for all $ A \subseteq S,
t \in T$.

Let us now interpret the KKT conditions at the \emph{OPT flow-rate} $(f^{*},R^{*})$. Suppose that
$f_{P}^*
> 0$  for $P \in \mathcal{P}_{s,t}$. Then due to complementary slackness, we have $\mu_{P}^* = 0$ and
consequently from equation (\ref{kkt-flow}) we get
$\sum_{e \in P} c_e^{'}(z_e^*) z_{e,t}^{'}(\bx_e^*) =
\lambda_{s,t}^*$
i.e. if there exists another path $Q \in \mathcal{P}_{s,t}$
such that $f_{Q}^* > 0$ then
$\sum_{e \in P} c_e^{'}(z_e^*) z_{e,t}^{'}(\bx_e^*) = \sum_{e \in
Q} c_e^{'}(z_e^*) z_{e,t}^{'}(\bx_e^*)$.

Now if we interpret
the quantity $\sum_{e \in P} c_e^{'}(z_e) z_{e,t}^{'}(\bx_e)$ as the
  \textit{differential cost} of the path $P$ associated with the flow-rate $(f,R)$ then this condition
implies that the differential cost of all the paths going from the same source to the
same terminal with positive flows at
\emph{OPT} is the same. It is quite intuitive for if it were not true
the objective function could be further decreased by moving some flow from
a higher differential cost path to a lower differential cost one without violating
feasibility conditions, and of course this should
not be possible at the optimum. Similarly, the differential cost
along a path with zero flow at OPT must have higher differential cost and indeed
this can be obtained as above by further noting that the dual
variables $\mu_P$'s are non-negative.
We note this property of the OPT flow-rate
in the following lemma.
\begin{lemma}
\label{mult-opt1}
Let  $(f^{*},R^{*})$  be an  OPT flow-rate for the instance $(G,c,d, \mathcal{R_{SW}})$.
Then, $\forall t \in T, ~ \forall s \in S$, $P, Q \in \mathcal{P}_{s,t}$ with $f_{P} > 0$ we have
\begin{align*}
\sum_{e \in P} c_e^{'}(z_e^{*}) z_{e,t}^{'}(\bx_e^{*}) \leq \sum_{e \in
Q} c_e^{'}(z_e^{*}) z_{e,t}^{'}(\bx_e^{*}). 
\end{align*}
\end{lemma}

The above lemma provides a simple and intuitive characterization of how the
flow allocations on various paths of same type (that is originating at
same source and ending at the same terminal) behave at the optimum
solution. Although such a
simple and intuitive characterization of the behavior of joint flow and
rate allocations at optimum is  not immediately clear, we can indeed
obtain three other simple and intuitive conditions that together with
Lemma \ref{mult-opt1}, are equivalent to the KKT conditions.
We establish this important characterization in the Theorem \ref{fourcondopt}.
First, we will show in the next three lemmas that these conditions are necessary
for optimality.
\begin{lemma}
\label{optcond4}
Let $(f,R)$ be an OPT flow-rate for the instance $(G,c,d,\mathcal{R_{SW}})$.
For $t \in T$, suppose that there exist $i,j \in S$ that satisfy the following property. If $A \subseteq S$, such that $i \in A$ and $\sum_{l \in A} R_{l,t} = H(X_A | X_{-A})$, then $j \in A$. For such $i$ and $j$ let $P \in \mathcal{P}_{i,t}, Q \in \mathcal{P}_{j,t}$ with $f_P > 0$. Then
\begin{align*}
 \sum_{e \in P} c_e^{'}(z_e) z_{e,t}^{'}(\bx_e) + d_i^{'}(y_i) y_{i,t}^{'}(\brho_i) 
 \leq \sum_{e \in Q} c_e^{'}(z_e) z_{e,t}^{'}(\bx_e)+ d_j^{'}(y_j) y_{j,t}^{'}(\brho_j).
\end{align*}
\end{lemma}
\proof
Since $(f,R)$ is an OPT flow-rate, it satisfies the KKT conditions for some
suitable choice of dual variables $\lambda_{i,t} \geq 0$, $\mu_P \geq 0$, $\nu_{A,t} \geq 0$.
Now, we are given that $ j \in A$ for all $A \subseteq S$ such that $i \in A$ and $\sum_{l \in A} R_{l,t} =H(X_{A}|X_{-A})$,
so if there is an $A \subseteq S$ such that $i \in A$ but $j \notin A$ then $\sum_{l \in A} R_{l,t} > H(X_{A}|X_{-A})$
and therefore by complementary slackness we get $\nu_{A,t} = 0$.
Further, from Equation \ref{kkt-flow-rate}, we have
\begin{eqnarray*}
d_i^{'}(y_i)
y_{i,t}^{'}(\brho_i) + \lambda_{i,t} &=& \sum_{A \subseteq
S:i \in A} \nu_{A,t} \\
&=& \sum_{A \subseteq
S:i \in A, j \in A} \nu_{A,t} \\
& & \text{~ (since $\sum_{A \subseteq
S:i \in A, j \notin A} \nu_{A,t} = 0$)}
\end{eqnarray*}
and
\begin{eqnarray*}
 d_j^{'}(y_j)
y_{j,t}^{'}(\brho_j) + \lambda_{j,t} &=& \sum_{A \subseteq
S:j \in A} \nu_{A,t} \\
&=& \sum_{A \subseteq
S:j \in A,i \in A} \nu_{A,t} 
 + \sum_{A \subseteq
S:j \in A, i \notin A} \nu_{A,t} \\
&\geq&  \sum_{A \subseteq
S:j \in A,i \in A} \nu_{A,t}\\
&=& d_i^{'}(y_i) y_{i,t}^{'}(\brho_i) + \lambda_{i,t}.
\end{eqnarray*}
Therefore we get,
\begin{equation*}
d_i^{'}(y_i) y_{i,t}^{'}(\brho_i) + \lambda_{i,t} \leq  d_j^{'}(y_j) y_{j,t}^{'}(\brho_j) + \lambda_{j,t}.
\end{equation*}
Furthermore, we are given that $f_P > 0$
which, using Equation \ref{kkt-flow} and complementary slackness condition $f_P\mu_P=0$, implies that
$\lambda_{i,t} = \sum_{e \in P} c_e^{'}(z_e) z_{e,t}^{'}(\bx_e) $ and
since $\mu_Q \geq 0$ we have $\sum_{e \in Q} c_e^{'}(z_e) z_{e,t}^{'}(\bx_e) \geq \lambda_{j,t}$.
Therefore,
\begin{align*}
& d_i^{'}(y_i) y_{i,t}^{'}(\brho_i) + \sum_{e \in P} c_e^{'}(z_e) z_{e,t}^{'}(\bx_e) 
 \leq  d_j^{'}(y_j) y_{j,t}^{'}(\brho_j)
+ \sum_{e \in Q} c_e^{'}(z_e) z_{e,t}^{'}(\bx_e).
\end{align*}
This concludes the proof. \endproof
\begin{lemma}
\label{optcond1}
Let $(f,R)$ be an OPT flow-rate for the instance
$(G,c,d,\mathcal{R_{SW}})$ wherein the functions $c_e$'s and $d_s$'s are
all strictly convex, then
$\forall t \in T, ~ ~ \forall s \in S$, we have
$\sum_{P \in \mathcal{P}_{s,t}} f_P = R_{s,t}$.
\end{lemma}
\proof
Let $\sum_{P \in \mathcal{P}_{s,t}} f_P > R_{s,t}$ then there is a $P \in P_{s,t}$ with $f_P > 0$.
Define a new feasible flow $\tilde{f}$ such that $\tilde{f}_Q= f_Q$ if $ Q \neq P$ and $\tilde{f}_P=f_P -\delta$
for some $ 0 < \delta < \min \{ f_P, \sum_{P \in \mathcal{P}_{s,t}} f_P - R_{s,t}\}$. Then,
\begin{eqnarray*}
\sum_{e \in E} c_e(\tilde{z}_e) &=& \sum_{e \in P} c_e(\tilde{z}_e) + \sum_{e \notin P} c_e(z_e) \\
& =& \sum_{e \in E} c_e(z_e) + \sum_{e \in P} \left(c_e(\tilde{z}_e) -  c_e(z_e)\right) \\
\end{eqnarray*}
\emph{Now, since the functions $c_e$ is non-decreasing as well as $z_e$ is non-decreasing in each co-ordinate, we get
$c_e(\tilde{z}_e) -  c_e(z_e) \leq 0$ for all $ e \in P$}.
Therefore,
\begin{eqnarray*}
\sum_{e \in E} c_e(\tilde{z}_e) &\leq & \sum_{e \in E} c_e(z_e)  ~ \implies \\
C(\tilde{f},R) &=&\sum_{e \in E} c_e(\tilde{z}_e) + \sum_{s \in S} d_s(y_s) \\
 &\leq &  \sum_{e \in E} c_e(z_e)  + \sum_{s \in S} d_s(y_s)\\
& =& C(f,R)
\end{eqnarray*}
which is a contradiction because $(f,R)$, due to strict convexity of
the function $C$, is the \emph{unique} OPT flow-rate. \endproof

\begin{lemma}
\label{optcond2}
Let $(f,R)$ be an OPT flow-rate for the instance $(G,c,d,\mathcal{R_{SW}})$ wherein the functions $c_e$'s and $d_s$'s are
all strictly convex, then
$\forall t \in T$, we have
$\sum_{s \in S} R_{s,t} = H(X_S).$
\end{lemma}
\proof
As $R$ is feasible,  $\forall t \in T$, $\bR_t \in  \mathcal{R_{SW}}$
and therefore, $ \sum_{s \in S} R_{s,t} \geq H(X_S)$.
Suppose $ \sum_{s \in S} R_{s,t} > H(X_S) $ for some $t \in T$,
then from Theorem \ref{thm:sum-rate-constraint} there exist
an $s \in S $, such that all (Slepian-Wolf) inequalities in which $R_{s,t}$
participates are loose.
Therefore, we can decrease this rate $R_{s,t}$
by a positive amount $r$ i.e. to $\tilde{R}_{s,t} = R_{s,t} - r$, without violating feasibility.
This means that
we can define a feasible rate
$\tilde{R}$ such that $\tilde{R}_{i,t} = R_{i,t}$ if $i \neq s$ and $\tilde{R}_{s,t} = R_{s,t} - r$ for some
$ r > 0$.
Now,
\begin{eqnarray*}
\sum_{i \in S} d_i(\tilde{y}_i) &=& \sum_{i \in S}d_i(y_i) + \left(d_s(\tilde{y}_s) - d_s(y_s)\right) \\
\end{eqnarray*}
\emph{Now, since $d_s$ is non-decreasing as well as $y_s$ is non-decreasing in each co-ordinate, we get
$d_s(\tilde{y}_s) \leq d_s(y_s)$.}
Therefore,
\begin{eqnarray*}
\sum_{i \in S} d_i(\tilde{y}_i) &\leq& \sum_{i \in S}d_i(y_i)  ~ \implies \\
C(f,\tilde{R}) &=&\sum_{e \in E} c_e(z_e) + \sum_{s \in S} d_s(\tilde{y}_s) \\
 &\leq&  \sum_{e \in E} c_e(z_e)  + \sum_{s \in S} d_s(y_s)\\
& =& C(f,R)
\end{eqnarray*}
which is a contradiction because $(f,R)$, due to strict convexity of
the function $C$, is the \emph{unique} OPT flow-rate. \endproof


\begin{theo}
\label{fourcondopt}
A feasible flow-rate $(f,R)$ for the instance $(G,c,d,
\mathcal{R_{SW}})$, which satisfies the following four conditions is
an OPT flow-rate for the instance $(G,c,d,\mathcal{R_{SW}})$.
Also, there is always an OPT flow-rate that satisfies these four
conditions.
Further, when the edge cost functions $c_e$ for all $e \in E$ and
the source cost functions $d_s$ for all $ s\in S$ are strictly convex,
that is when the optimization problem \emph{(NIF-CP)} is strictly
convex, these conditions are also necessary for optimality.
\begin{enumerate}
\item  $\forall t \in T, ~ ~ \forall s \in S$, we have
\begin{equation*}
\sum_{P \in \mathcal{P}_{s,t}} f_P = R_{s,t}.
\end{equation*}

\item  $\forall t \in T$, we have
\begin{equation*}
\sum_{s \in S} R_{s,t} = H(X_S).
\end{equation*}

\item $\forall t \in T, ~ \forall s \in S$,
$P, Q \in \mathcal{P}_{s,t}$ with $f_{P} > 0$,
\begin{equation*}
\sum_{e \in P} c_e^{'}(z_e) z_{e,t}^{'}(\bx_e) \leq \sum_{e \in
Q} c_e^{'}(z_e) z_{e,t}^{'}(\bx_e).
\end{equation*}

\item  For $t \in T$, suppose that there exist $i,j \in S$ that satisfy
the following property. If $A \subseteq S$, such that $i \in A$ and
$\sum_{l \in A} R_{l,t} = H(X_A | X_{-A})$, then $j \in A$. For such
$i$ and $j$ let $P \in \mathcal{P}_{i,t}, Q \in \mathcal{P}_{j,t}$
with $f_P > 0$. Then
\begin{align*}
& \sum_{e \in P} c_e^{'}(z_e) z_{e,t}^{'}(\bx_e) + d_i^{'}(y_i)
y_{i,t}^{'}(\brho_i) 
 \leq \sum_{e \in
Q} c_e^{'}(z_e) z_{e,t}^{'}(\bx_e)+ d_j^{'}(y_j) y_{j,t}^{'}(\brho_j).
\end{align*}
\end{enumerate}
\end{theo}
\proof
We prove that the above four conditions imply optimality of
$(f,R)$.
Our assumptions guarantee that the optimization problem ({\em NIF-CP}) for the instance
$(G,c,d, \mathcal{R_{SW}})$ is convex and since all the
feasibility constraints are linear, 
strong duality holds \cite{boyd_van}. This implies
that the KKT conditions
are necessary and sufficient for optimality.
We show 
that a feasible flow-rate $(f,
R)$ with the above four properties satisfies the KKT conditions for
the instance $(G,c,d, \mathcal{R_{SW}})$
for a suitable choice of the dual variables given below. 

\noindent \textbf{Choosing $\lambda_{i,t}$}'s:
\begin{equation*}
\lambda_{i,t} := \min_{P \in \mathcal{P}_{i,t}} ~ \sum_{e \in P} c_e^{'}(z_e) z_{e,t}^{'}(\bx_e).
\end{equation*}
Note that, using \textbf{Condition 3}, for $i\in S$, if there exist a $P_i \in \mathcal{P}_{i,t}$
such that $f_{P_i} > 0$ then we have
$$\lambda_{i,t} = \sum_{e \in P_i}  c_e^{'}(z_e)z_{e,t}^{'}(\bx_e).$$

\noindent \textbf{Choosing $\mu_P$}'s:
For $P \in P_{i,t}$ take
\begin{equation*}
\mu_P := \sum_{e \in P}  c_e^{'}(z_e) z_{e,t}^{'}(\bx_e) -
\lambda_{i,t}.
\end{equation*}

\noindent \textbf{Choosing $\nu_{A,t}$}'s:
Let
$$h_{i,t} := d_i^{'}(y_i) y_{i,t}^{'}(\brho_i) +
\lambda_{i,t}.$$
Let $\pi$ denote a permutation such that
$0 \leq h_{\pi(1),t} \leq h_{\pi(2),t} \leq \dots h_{\pi(N_S),t}.$
Now take
%
\begin{align*}
\nu_{A,t} = \left\{ \begin{array}{l}
h_{\pi(1),t} ~ \textrm{if~} A = \{ \pi(1), \pi(2),\dots, \pi(N_S)\}\\
h_{\pi(i),t}- h_{\pi(i-1),t} ~ \textrm{if~} A = \{ \pi(i), \dots, \pi(N_S)\} \\
~~~~~~~~~~~~~~~~~~~~~~ \text{and~} 2 \leq i \leq N_S \\
0  ~~~ \textrm{ otherwise.}
\end{array}
\right.
\end{align*}

Now, with the above choice of dual variables we will check all the KKT
conditions one by one.

\noindent \textbf{Dual Feasibility:}
\begin{itemize}
\item $\lambda_{i,t} \geq 0$ as $c_e$ and $z_e$ are non-decreasing functions
  i.e. $c_e^{'}(z_e) \geq 0$ and $ z_{e,t}^{'}(\bx_e)  \geq 0$.

\item $\mu_{P} \geq 0$ by the definition because $\lambda_{i,t} \leq
  \sum_{e \in P}  c_e^{'}(z_e) z_{e,t}^{'}(\bx_e) ~ \forall P \in
  P_{i,t}$.

\item $\nu_{A,t} \geq 0$ by definition.
\end{itemize}

\noindent \textbf{KKT Conditions as per equation \ref{kkt-flow}:}

\begin{align*}
\frac{\partial L}{\partial f_{P}} &= \sum_{e \in P}
c_e^{'}(z_e) z_{e,t}^{'}(\bx_e) - \lambda_{i,t} - \mu_{P} \\
&= \sum_{e \in P}
c_e^{'}(z_e) z_{e,t}^{'}(\bx_e) - \lambda_{i,t}
 - \left(\sum_{e \in P}  c_e^{'}(z_e) z_{e,t}^{'}(\bx_e) -
\lambda_{i,t}\right) \\
& = 0.
\end{align*}

\noindent \textbf{KKT Conditions as per equation \ref{kkt-flow-rate}:}

\begin{align*}
\frac{\partial L}{\partial R_{\pi(i),t}} &= d_{\pi(i)}^{'}(y_{\pi(i)})
y_{\pi(i),t}^{'}(\brho_{\pi(i)}) + \lambda_{\pi(i),t}
- \sum_{A \subseteq S:\pi(i) \in A} \nu_{A,t} \\
&= h_{\pi(i),t} -  \sum_{A \subseteq S:\pi(i) \in A} \nu_{A,t} \\
&= h_{\pi(i),t} - \sum_{j \in \{1,2,\dots,i\}}
\nu_{\{\pi(j), \pi(j+1),\dots, \pi(N_S)\},t}\\
&= h_{\pi(i),t} - \left[ h_{\pi(1),t}+ (h_{\pi(2),t} -h_{\pi(1),t}) \right. \\
& ~~~~~~~~~~ \left.  +(h_{\pi(3),t}
  -h_{\pi(2),t})+ \dots + (h_{\pi(i),t} -h_{\pi(i-1),t})\right] \\
&= h_{\pi(i),t} - h_{\pi(i),t}
= 0.
\end{align*}

\noindent \textbf{Complementary Slackness Conditions:}
\begin{itemize}
\item $\mu_{P} f_P = 0$ for all $P \in \mathcal{P}$.

Let $P \in \mathcal{P}_{i,t}$ and $f_P > 0$ then using \textbf{Condition $3$} and
definition  of $\lambda_{i,t}$ we get $$\sum_{e \in P} c_e^{'}(z_e)
z_{e,t}^{'}(\bx_e) =\lambda_{i,t}$$ and therefore, $$\mu_P = \sum_{e \in P} c_e^{'}(z_e)
z_{e,t}^{'}(\bx_e) -\lambda_{i,t} = 0.$$

\item $\lambda_{s,t} (R_{s,t} - \sum_{P \in \mathcal{P}_{s,t}} f_P) =
  0$ for all $ s \in S,
t \in T$.

This follows from the \textbf{Condition $1$}.

\item  $\nu_{A,t} \left(H(X_A |X_{A^c}) - \sum_{i \in A} R_{i,t}\right)  = 0$ for all $ A \subseteq S,
t \in T$.

Note that $\nu_{A,t} = 0$ except for $A = \{\pi(i), \pi(i+1), \dots, \pi(N_S)\}, \textrm{~for~} i = 1, 2, \dots, N_S$.
Therefore the only condition that needs to be checked is that if
\newline 
$ \sum_{j=i}^{N_S} R_{\pi(j),t} >
 H(X_{\pi(i)}, X_{\pi(i+1)}, \dots, X_{\pi(N_S)} | X_{\pi(i-1)}, \dots, X_{\pi(1)}),$
then $h_{\pi(i),t} - h_{\pi(i-1),t} = 0$.
\end{itemize}



Towards this end let $j \in  \{ \pi(i), \pi(i+1), \dots, \pi(N_S)\}$, and let $A_j$ be the minimum
cardinality set such that $j \in A_j$ and $\sum_{l
\in A_j} R_{l,t} = H(X_{A_j}|X_{-A_{j}})$ i.e.
\begin{equation*}
A_j = \arg \min_{A \subseteq S: j \in A,\sum_{l
\in A} R_{l,t} = H(X_{A}|X_{-A}) }  ~ |A|.
\end{equation*}
 Such a set $A_j$
always exists because from \textbf{Condition $2$} we have
$\sum_{l=1}^{N_S} R_{l,t} = H(X_1, \dots, X_{N_S})$ and therefore the
set $\left\{A \subseteq S: j \in A,\sum_{l \in A} R_{l,t} =
  H(X_{A}|X_{-A}) \right\}$ is not empty.

We claim that
there exists a $j^* \in \{ \pi(i), \pi(i+1), \dots, \pi(N_S)\}$ such that $A_{j^*} \cap
\{\pi(1), \pi(2),\dots, \pi(i-1)\}$ is not empty. If this is not true then clearly
we have $ \cup_{j=\pi(i)}^{\pi(N_S)} A_j = \{\pi(i), \pi(i+1),\dots, \pi(N_S)\}$ and using the supermodularity property of conditional entropy
(ref. \textbf{Lemma \ref{int-uni-rate-tight}}),
we obtain
\begin{eqnarray*}
\sum_{j= \pi(i)}^{\pi(N_S)}
R_{j,t} = 
 H(X_{\pi(i)}, X_{\pi(i+1)}, \dots, X_{\pi(N_S)}| X_{\pi(i-1)}, \dots, X_{\pi(1)}),
\end{eqnarray*}
which is a contradiction, therefore we must have such a $j^* \in  \{
\pi(i), \pi(i+1), \dots, \pi(N_S)\}$ such that $A_{j^*} \cap \{\pi(1), \pi(2),\dots, \pi(i-1)\}$ is not
empty. 

Next, we show that there exists a source $k \in \{\pi(1), \pi(2),\dots, \pi(i-1)\}$
such that if $j^* \in A$ and $\sum_{l \in A} R_{l,t} = H(X_A |X_{-A})$, then $k \in A$.
Towards this end suppose that there exist subsets $S_1$ and $S_2$ of $S$ such that $ j^* \in S_1 \cap S_2$ and
$\sum_{l \in S_1} R_{l,t} = H(X_{S_1}|X_{-S_{1}})$ and $\sum_{l \in S_2}  R_{l,t} = H(X_{S_2}|X_{-S_{2}})$,
then using the supermodularity property of conditional entropy
we can show that
rate inequality involving $ S_1 \cap S_2$ is
also tight ( \textbf{Lemma \ref{int-uni-rate-tight}}) ~~ i.e.
$\sum_{l \in  S_1 \cap S_2 } R_{l,t} = H(X_{ S_1 \cap S_2}|X_{- (S_1 \cap S_2)})$.
This implies that $A_{j^*}$, being of minimum cardinality, is the intersection of
all sets that have $j^*$ as a member on which the rate inequality is tight i.e.
\begin{equation*}
A_{j^*} = \bigcap_{A \subseteq S} ~ \{ A : j^* \in A, \sum_{l \in A} R_{l,t} =
H(X_{A}|X_{-A}) \}.
\end{equation*}
Moreover note that $A_{j^*}$ is not a singleton set since $A_{j^*} \cap  \{\pi(1), \pi(2),\dots, \pi(i-1)\} \neq \phi$.
Therefore there exists a $k \in A_{j^*}$ such that $k \neq j^*$. By our above arguments this implies that if $A \subseteq S$ is such that
$j^* \in A$ and $\sum_{l \in A} R_{l,t} =H(X_{A}|X_{-A})$ then $k \in A$.

Clearly, $R_{j^*,t} > H(X_{j^*}|X_{-j^*})$ as $k$ does not participate in this
rate inequality. Therefore, $R_{j^*,t} > 0$ which implies that there exists a $P \in
\mathcal{P}_{j^*,t}$ with $f_P > 0$, therefore using
\textbf{Condition $3$} and the definition of $\lambda_{j^*,t}$ we have
$\sum_{e \in P} c_e^{'}(z_e) z_{e,t}^{'}(\bx_e) = \lambda_{j^*,t}$.
Also, by the definition of  $\lambda_{k,t}$ there is a $Q \in
\mathcal{P}_{k,t}$ such that $\sum_{e \in
Q} c_e^{'}(z_e) z_{e,t}^{'}(\bx_e) = \lambda_{k,t}$.

Now using \textbf{Condition $4$}, we get
\begin{align*}
\sum_{e \in P} c_e^{'}(z_e) z_{e,t}^{'}(\bx_e)  +
d_{j^*}^{'}(y_{j^*}) y_{j^*,t}^{'}(\brho_{j^*}) 
 \leq \sum_{e \in
Q} c_e^{'}(z_e) z_{e,t}^{'}(\bx_e)+ d_k^{'}(y_k) y_{k,t}^{'}(\brho_k)
~~ \forall Q \in \mathcal{P}_{k,t}
\end{align*}
which implies that
\begin{equation*}
\lambda_{j^*,t}  + d_{j^*}^{'}(y_{j^*}) y_{{j^*},t}^{'}(\brho_{j^*}) \leq \lambda_{k,t} + d_k^{'}(y_k) y_{k,t}^{'}(\brho_k)
\end{equation*}
and therefore we get $h_{j^*,t} \leq h_{k,t}$. Now note that $k \in \{\pi(1), \pi(2), \dots, \pi(i-1)\}$ while $j^* \in \{\pi(i), \dots, \pi(N_S)\}$.
This implies in turn that $h_{\pi(i),t} \leq h_{j^*,t} \leq h_{k,t}$. But, we know that $h_{k,t}
\leq h_{\pi(i-1),t}$ i.e. $h_{\pi(i),t} - h_{\pi(i-1),t} \leq 0$ but
we already have $h_{\pi(i),t} - h_{\pi(i-1),t} \geq 0$ and hence $h_{\pi(i),t} - h_{\pi(i-1),t} = 0$.

This establishes that the four conditions are sufficient for optimality. Further, as per Lemmas \ref{mult-opt1}, \ref{optcond4},
\ref{optcond1}, \ref{optcond2}, under strict convexity conditions, these conditions are
necessary too.
\endproof

\begin{co}
\label{optnc}
If the sources are independent (i.e. $\mathcal{R_{SW}} \in
\mathcal{M}_{ind}$), there is a feasible flow-rate for instance
$(G,c,d,\mathcal{R_{SW}})$ that is an
OPT flow-rate for both the instances $(G,c,d,\mathcal{R_{SW}})$ and $(G,\tilde{c},\tilde{d},\mathcal{R_{SW}})$, where
$\tilde{c}_e(x)=\alpha c_e(x)$ for constant $\alpha > 0$, and
$\tilde{d}_s$ is \emph{any} convex, differentiable, positive and
non-decreasing function.
Further, this OPT flow-rate satisfies the four conditions in Theorem
\ref{fourcondopt} for both the instances  $(G,c,d,\mathcal{R_{SW}})$
and $(G,\tilde{c},\tilde{d},\mathcal{R_{SW}})$.
\end{co}

\proof The idea is that when the
sources are independent, Condition
(2) in Theorem \ref{fourcondopt} implies that $R_{s,t}= H(X_s)$ for
all $s \in S, t \in T$,
and therefore, there is no pair $(i,j)$ such that
$j$ participates in all tight rate inequalities involving $i$ and
consequently it is not required to check Condition (4). For the sake of completeness
the proof follows.

Let $(f,R)$ be an OPT flow-rate for $(G,c,d,\mathcal{R_{SW}})$
satisfying the four conditions in Theorem \ref{fourcondopt}. Note that
such an OPT flow-rate always exists as per Theorem \ref{fourcondopt}.
Since the sources are independent the rate inequalities constraints becomes
\begin{equation*}
\sum_{i \in A} R_{i,t} \geq H(X_A) ~ \textrm{ for all $ A \subseteq S, t \in T$.}
\end{equation*}
Therefore, using Condition (2) in Theorem \ref{fourcondopt}, we obtain
\begin{equation*}
R_{s,t}= H(X_s) ~ \textrm{ for all $s \in S, t \in T$.}
\end{equation*}

Now we will show that $(f,R)$ is also an OPT flow-rate for the instance  $(G,\tilde{c},\tilde{d},\mathcal{R_{SW}})$
by showing that it satisfies the four conditions in Theorem \ref{fourcondopt} for instance  $(G,\tilde{c},\tilde{d},\mathcal{R_{SW}})$.
Note that Conditions (1) and (2) are easily satisfied by $(f,R)$ as they do not depend on particular cost functions.
Further,
$$\sum_{e \in P} \tilde{c}_e^{'}(z_e) z_{e,t}^{'}(\bx_e) = \alpha \sum_{e \in P} c_e^{'}(z_e) z_{e,t}^{'}(\bx_e), $$
therefore condition $$\sum_{e \in P} \tilde{c}_e^{'}(z_e) z_{e,t}^{'}(\bx_e) \leq \sum_{e \in Q} \tilde{c}_e^{'}(z_e) z_{e,t}^{'}(\bx_e)$$
is equivalent to $$\sum_{e \in P} c_e^{'}(z_e) z_{e,t}^{'}(\bx_e)  \leq \sum_{e \in Q} c_e^{'}(z_e) z_{e,t}^{'}(\bx_e),$$
therefore condition (3) is also satisfied.
For the condition (4), let us first note that as discussed above $R_{s,t}=H(X_s)$ for all $ s\in S, t \in T$. This implies that
there is no pair $(i,j) \in S \times S$ satisfying the promise in condition (4) i.e. there is no pair $(i,j)$ such that
$j$ participates in all tight rate inequalities involving $i$
(simply because $j$ does not participate in the tight rate inequality $R_{i,t}=H(X_i)$).
Thus, $(f,R)$ satisfies all the $4$ conditions in Theorem \ref{fourcondopt} for the instance
$(G,\tilde{c},\tilde{d},\mathcal{R_{SW}})$ and hence is an OPT flow-rate for  $(G,\tilde{c},\tilde{d},\mathcal{R_{SW}})$.
\endproof

\section{The Flows and Rates at Nash Equilibrium }
\label{nash-flow-rate-prop-sec}
In this section, we study the properties of a Nash flow-rate whenever
the individual optimization problem (i.e. the best response problem) of each terminal is convex, that
is whenever Nash equilibrium can be considered as an appropriate solution
concept for the Distributed Compression Game when viewed through the
algorithmic lens.
Therefore, throughout this section, we assume that the edge cost
splitting mechanism $\Psi$, as well as, the source cost splitting
mechanism $\Phi$ are such that the functions $C^{(t)}$, for all $t \in
T$, are convex.  By considering the best response problem of each
terminal, and an approach essentially the same as in the Section
\ref{opt-flow-rate-problem} for characterizing OPT flow-rate, we
can obtain the following Theorem \ref{mult-nash1} for characterizing
Nash flow-rate.

\begin{theo}
\label{mult-nash1}
Consider an instance $(G,c,d,\mathcal{R_{SW}}, \Psi,\Phi)$ where $C^{(t)}$ is
convex for all $ t \in T$.
A feasible flow-rate $(f,R)$ for the instance
$(G,c,d,\mathcal{R_{SW}})$, which satisfies the following four conditions is a Nash flow-rate for $(G,c,d,\mathcal{R_{SW}}, \Psi,
\Phi)$. Further, when $C^{(t)}$ is
strictly convex for all $ t \in T$, these conditions are also necessary.
\begin{itemize}
\item[(1)] $\forall t \in T, ~ ~ \forall s \in S$, we have
\begin{equation*}
\sum_{P \in \mathcal{P}_{s,t}} f_P = R_{s,t}.
\end{equation*}

\item[(2)] $\forall t \in T$, we have
\begin{equation*}
\sum_{s \in S} R_{s,t} = H(X_S).
\end{equation*}

\item[(3)]  $\forall t \in T, ~ \forall s \in S$,
$P, Q \in \mathcal{P}_{s,t}$ with $f_{P} > 0$,
\begin{equation*}
\frac{\partial C_E^{(t)}(f)}{\partial f_P} \leq \frac{\partial C_E^{(t)}(f)}{\partial f_Q}.
\end{equation*}

\item[(4)] For $t \in T $, let $j \in S$ participates in \textbf{all tight} rate inequalities involving
$i \in S$ (i.e. if $A \subseteq S$, such that $i \in A$ and $\sum_{l \in A} R_{l,t} = H(X_A | X_{-A})$, then $j \in A$) and
let $P \in \mathcal{P}_{i,t}, Q \in \mathcal{P}_{j,t}$ with $f_P > 0$ then we have
\begin{equation*}
\frac{\partial C_E^{(t)}(f)}{\partial f_P}  + \frac{\partial C_S^{(t)}(R)}{\partial R_{i,t}}  \leq \frac{\partial C_E^{(t)}(f)}{\partial f_Q} + \frac{\partial C_S^{(t)}(R)}{\partial R_{j,t}}.
\end{equation*}
\end{itemize}
\end{theo}

Further, under similar convexity conditions, we can also show that a Nash flow-rate always exists for the
Distributed Compression Game. This is done via first compactifying the
strategy sets $A_t$'s to obtain a restricted game where existence of a
Nash equilibrium follows from the standard fixed point
theorems \cite{ORgamebook}. Then, by utilizing the monotonically
non-decreasing properties of various cost functions, it is argued that a
Nash equilibrium of the restricted game is also a Nash flow-rate for
our \emph{Distributed Compression Game} thereby proving the existence of a Nash
flow-rate for \emph{Distributed Compression Game}.

The Theorem  \ref{katukani-nash1} in the following is a very standard and popular result on the existence of Nash equilibrium
and we adopt it from the book by Osborne and Rubinstein \cite{ORgamebook}.

\begin{theo}
\label{katukani-nash1}
The strategic game $\langle \mathcal{N}, \left(A_i\right), \left(\succeq_i\right) \rangle$ has a Nash equilibrium if
for all $i \in \mathcal{N}$, the following conditions hold.
\begin{itemize}
\item[a)] The set $A_i$ of actions of player $i$ is a nonempty compact convex subset of a Euclidean space.

\item[b)] The preference relation $\succeq_i$ is continuous and quasi-concave on $A_i$. A preference relation
 $\succeq_i$ on $\mathcal{A}$ is said to be quasi-concave on $A_i$ if for
 every $a \in \mathcal{A}$ the set $\left\{\tilde{a}_i \in
   A_i:(a_{-i},\tilde{a}_i) \succeq_i a \right\}$ is convex. A preference relation
 $\succeq_i$ on $\mathcal{A}$ is said to be continuous if $a
 \succeq_i b$ whenever there
 are sequences $\{a^k\}$ and $\{b^k\}$ with $a^k, b^k \in \mathcal{A}$
 and $a^k \succeq_i b^k$ for all $k$ such that  $\{a^k\}$ and $\{b^k\}$
 converge to $a$ and $b$ respectively.
\end{itemize}
\end{theo}

Now, let us consider an instance $(G,c,d,\mathcal{R_{SW}},\Psi,\Phi)$ of the Distributed Compression Game, where $C^{(t)}$ is
convex for all $ t \in T$.

The action set of the terminal $t \in T$ is
\begin{equation}
A_t = \left\{  (f_t,\bR_t): \begin{array}{l} f_P  \geq 0 ~~ \forall P \in \mathcal{P}_t, \\
 \sum_{P \in \mathcal{P}_{s,t}} f_P  \geq R_{s,t} ~~  \forall s \in S, \\
 \bR_t  \in \mathcal{R_{SW}}
\end{array}
\right\}.
\end{equation}

Clearly this is a nonempty convex subset of an Euclidean Space, but it is not compact.

Let us consider a game with a restricted set of strategies denoted $\tilde{A}_t$'s
 as follows and let us call this new game as the \textbf{restricted game}
for the instance $(G,c,d,\mathcal{R_{SW}},\Psi,\Phi)$.

\begin{equation}
\tilde{A}_t = \left\{  (f_t,\bR_t): \begin{array}{l} f_P  \geq 0 ~~ \forall P \in \mathcal{P}_t, \\
 \sum_{P \in \mathcal{P}_{s,t}} f_P  \geq R_{s,t} ~~  \forall s \in S, \\
 \bR_t  \in \mathcal{R_{SW}}, \\
f_P \leq H(X_S) ~~ \forall P \in \mathcal{P}_t ,\\
R_{s,t} \leq H(X_S) ~~  \forall s \in S
\end{array}
\right\}.
\end{equation}

Now the set $\tilde{A}_t$ becomes compact as it is a closed and bounded subset of an Euclidean space, and therefore
$\tilde{A}_t$ satisfies the requirement $(a)$ of the Theorem \ref{katukani-nash1}.

Since  players' cost functions $C^{(t)}$ are
convex and continuous for all $ t \in T$, the condition $(b)$ in the Theorem \ref{katukani-nash1}
is also satisfied and we obtain the following result.

\begin{lemma}
\label{existlm1}
The restricted game for the instance $(G,c,d,\mathcal{R_{SW}},\Psi,\Phi)$, where $C^{(t)}$ is
convex for all $ t \in T$, admits a Nash equilibrium.
\end{lemma}

Now we claim that every Nash equilibrium of the restricted game is also a Nash equilibrium for the original game and
that will imply the existence of a Nash flow-rate for the original game.

\begin{lemma}
\label{existlm2}
Every Nash equilibrium of the restricted game for the instance $(G,c,d,\mathcal{R_{SW}},\Psi,\Phi)$, where $C^{(t)}$ is
convex for all $ t \in T$, is also a Nash flow-rate
for the instance $(G,c,d,\mathcal{R_{SW}},\Psi,\Phi)$.
\end{lemma}

\proof
Let $(f,R)$ be a Nash equilibrium of the restricted game for the instance $(G,c,d,\mathcal{R_{SW}},\Psi,\Phi)$. Then, for all $t$
we have
\begin{equation*}
C^{(t)}(f,R) \leq C^{(t)}(f_{-t},\bR_{-t},\tilde{f}_t,\tilde{\bR}_t)
\end{equation*}
for all $\tilde{f}_t,\tilde{\bR}_t$ feasible for the restricted game i.e. coming from the restricted strategy set $\tilde{A}_t$.

Now let $(\tilde{f}_t,\tilde{\bR}_t) \in A_t \setminus \tilde{A}_t$ i.e.  $\tilde{f}_t,\tilde{\bR}_t$ is feasible for the original
game but not feasible for the restricted game. For ease of notation, let us define the following quantities.
\begin{eqnarray*}
S_{1,t} = \left\{ s \in S: \tilde{R}_{s,t} > H(X_S)\right\} ~~,~~~ S_{2,t} = S \setminus S_{1,t}\\
\bR_{t}^{'} = \left\{  R_{s,t}^{'}:=H(X_S) | s \in S_{1,t} \right\} \\
\mathcal{P}_t^1 = \left\{ P \in \mathcal{P}_t : \tilde{f}_P >  H(X_S) \right\} ~~,~~~ \mathcal{P}_t^2 = \mathcal{P}_t \setminus \mathcal{P}_t^1\\
f_{t}^{'} = \left\{  f_{P}^{'}:=H(X_S) | P \in \mathcal{P}_t^1 \right\} \\
\end{eqnarray*}
Note that in defining $\bR_{t}^{'}$ and  $f_{t}^{'}$ we have projected all the flows and rates violating the feasibility
for the restricted game to their boundary values and therefore the strategy
$(f_{t}^{'}, \{\tilde{f}_P:P \in \mathcal{P}_t^2 \}, \bR_{t}^{'}, \{\tilde{R}_{s,t}: s \in S_{2,t}\}) \in \tilde{A}_t$ i.e. it is
feasible for the restricted game.

Now,
\begin{eqnarray*}
 C^{(t)}(f_{-t}, \bR_{-t},\tilde{f}_t, \tilde{\bR}_t) 
   && \geq C^{(t)}(f_{-t}, \bR_{-t},\tilde{f}_t, \bR_{t}^{'}, \{\tilde{R}_{s,t}: s \in S_{2,t}\}) \\
   &&\geq C^{(t)}(f_{-t}, \bR_{-t},f_{t}^{'}, \{\tilde{f}_P:P \in \mathcal{P}_t^2 \}, \bR_{t}^{'}, \{\tilde{R}_{s,t}: s \in S_{2,t}\}) \\
\end{eqnarray*}
and since $(f,R)$ is a Nash equilibrium for the restricted game and
 $(f_{t}^{'}, \{\tilde{f}_P:P \in \mathcal{P}_t^2 \}, \bR_{t}^{'}, \{\tilde{R}_{s,t}: s \in S_{2,t}\})$ is
feasible for the restricted game we have
\begin{eqnarray*}
 C^{(t)}(f,R) 
&\leq& C^{(t)}(f_{-t}, \bR_{-t},f_{t}^{'}, \{\tilde{f}_P:P \in \mathcal{P}_t^2 \}, \bR_{t}^{'}, \{\tilde{R}_{s,t}: s \in S_{2,t}\})\\
   & \leq & C^{(t)}(f_{-t}, \bR_{-t},\tilde{f}_t, \tilde{\bR}_t)
\end{eqnarray*}
and therefore $C^{(t)}(f,R) \leq C^{(t)}(f_{-t}, \bR_{-t},\tilde{f}_t, \tilde{\bR}_t) $ for all $(\tilde{f}_t, \tilde{\bR}_t) \in A_t$ implying
that $(f,R)$ is a Nash equilibrium of the original game meaning  $(f,R)$ is a Nash flow-rate for the instance $(G,c,d,\mathcal{R_{SW}},\Psi,\Phi)$ \endproof

Combining the Lemmas \ref{existlm1} and \ref{existlm2} we obtain the following theorem.

\begin{theo}
\label{existth1}
An instance $(G,c,d,\mathcal{R_{SW}},\Psi, \Phi)$, where $C^{(t)}$ is
convex for all $ t \in T$,  admits a Nash flow-rate.
\end{theo}

\section{Wardrop Flow-Rate and the Price of Anarchy}
\label{waldroppoasec}
In this section, we investigate the inefficiency brought forth by the selfish behavior of
terminals.
First, we will show that the Wardrop equilibrium is a socially optimal solution for a
different set of (related) cost functions.
Using this, we will construct explicit examples that demonstrate that
the POA $> 1$ and determine near-tight upper bounds on the POA as
well. We start out with the characterization of Wardrop flow-rate.
\begin{theo}
\label{waldropopteq}
Let  $z_e(\bx_e) =\left(\sum_{t \in T} x_{e,t}^n\right)^{\frac{1}{n}}, \Psi_{e,t}(\bx_e)  = \frac{ x_{e,t}^n}{ \left(\sum_{j \in T} x_{e,j}^n\right)}$
and $  \Phi_{s,t}(\brho_s)= \frac{1}{N_T}$.
 A Wardrop flow-rate for $(G,c,d,\mathcal{R_{SW}}, \Psi, \Phi)$ is an OPT flow-rate for $(G,\tilde{c},d,\mathcal{R_{SW}})$,
where $\tilde{c}_e(x) = N_T ~ \int \frac{c_e(x)}{x} dx$. Further, when the edge cost functions $c_e$ for all $e \in E$ and
the source cost functions $d_s$ for all $ s\in S$ are strictly convex,
an OPT flow-rate for $(G, c,d,\mathcal{R_{SW}})$ is also a Wardrop
flow-rate for $(G,\hat{c},d,\mathcal{R_{SW}}, \Psi, \Phi)$, where
$\hat{c}_e(x)=\frac{1}{N_T} x c_e^{'}(x)$.
\end{theo}
\proof
We will show that the definition of a Wardrop flow-rate for instance $(G,c,d,\mathcal{R_{SW}}, \Psi, \Phi)$ exactly corresponds
to the four conditions for the instance $(G,\tilde{c},d,\mathcal{R_{SW}})$ in Theorem \ref{fourcondopt}.

We have, 
\begin{align*}
z_{e,t}^{'}(\bx_e) = \frac{1}{n} \left(\sum_{j \in T}
  x_{e,j}^n\right)^{\frac{1}{n} - 1} n x_{e,t}^{n-1}
= \frac{z_e}{x_{e,t}} \frac{x_{e,t}^n}{\sum_{j \in T} x_{e,j}^n}.
\end{align*}
Therefore,
\begin{align*}
  C_P(f) & =\sum_{e \in P}  c_e(z_e)  \frac{ x_{e,t}^{n-1}}{ \left(\sum_{j \in T} x_{e,j}^n\right)}\\
 &= \sum_{e \in P}  c_e(z_e) \frac{z_{e,t}^{'}(x_e)}{z_e}\\
& = \frac{1}{N_T} \sum_{e \in P}  \tilde{c}_e^{'}(z_e) z_{e,t}^{'}(x_e)
\end{align*}
where the last equality follows from the fact that
$$\tilde{c}_e(x) = N_T \int \frac{c_e(x)}{ x} dx ~  \Longrightarrow \tilde{c}_e^{'}(x) =  N_T \frac{c_e(x)}{x}.$$
 Also,
$$C_S^{(t)}(R)  = \frac{1}{N_T} \sum_{i \in S} d_i(y_i), \implies$$
 $$\frac{\partial C_S^{(t)}(R)}{\partial R_{i,t}}   = \frac{1}{N_T} d_{i}^{'}(y_i) y_{i,t}^{'}(\brho_i).$$ Therefore,
\begin{align*}
C_{P}(f) + \frac{\partial C_S^{(t)}(R)}{\partial R_{i,t}}
 = \frac{1}{N_T} \left[ \sum_{e \in P}  \tilde{c}_e^{'}(z_e) z_{e,t}^{'}(\bx_e)+
d_{i}^{'}(y_i) y_{i,t}^{'}(\brho_i)\right].
\end{align*}
The result follows from the equivalence of conditions coming from
Definition \ref{waldropdef} and Theorem \ref{fourcondopt}. \endproof

\begin{figure}
\centering
\scalebox{1.0}{\includegraphics{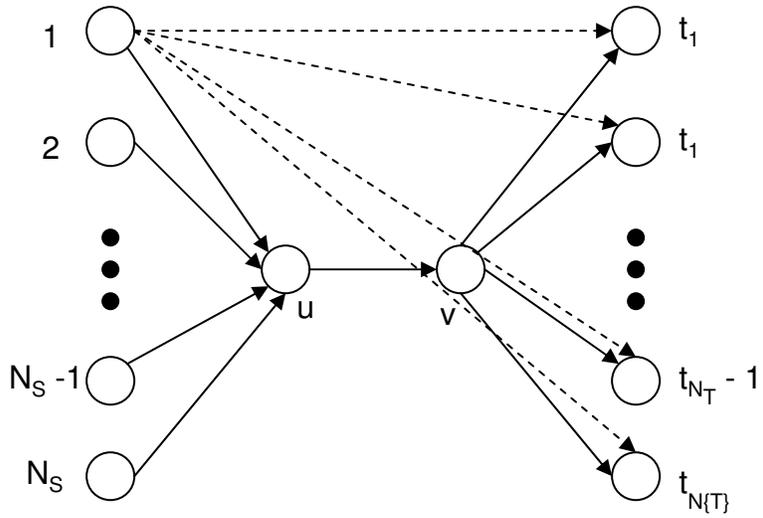}}
\caption{Example of a network where POA is linear in $N_T$.}
\label{poafig01}
\end{figure}

\begin{figure}
\centering
\scalebox{1.0}{\includegraphics{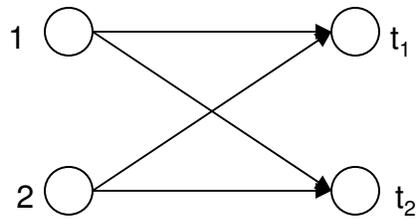}}
\caption{Classical Slepian-Wolf network with appropriate costs also has POA $>$ 1.}
\label{poafig02}
\end{figure}

In contrast with the result of \cite{BhadraSG06} that holds for a single source with the edge cost splitting mechanism used above, from Theorem \ref{waldropopteq}, we can note that for most reasonable
cost splitting mechanisms, the POA will not equal one for all monomial edge cost functions.
We construct explicit examples for POA $> 1$ in the Figures \ref{poafig01} and \ref{poafig02}.
The example in Figure \ref{poafig01} is near tight as will be evident from
an upper bound on POA derived in Theorem \ref{poa-ubound}.

It is interesting to note that in the case when sources are independent, in the Wardrop
or OPT solutions, the rates requested at various sources will equal their respective lower bounds
(i.e. their entropies). Therefore, the cost term corresponding to
the sources will be fixed, and one only needs to find flows that minimize the edge costs. In this situation, it is not hard to see that the POA will again equal one
for \textit{ all} monomial edge cost functions. i.e. {\em it is the correlation among the sources
that is responsible for bringing more anarchy}. We formalize this below.

Let $\mathcal{C}_k =\{ c: c_e(x)=a_e x^k, ~  a_e > 0 , \forall e \in E\}$ be the set of edge cost functions
where all edge cost functions are monomial of the same degree $k$ possibly with different coefficients,
and $\mathcal{C}_{mon} = \cup_{ k \geq 1} \mathcal{C}_k$.
Similarly, $\mathcal{D}_k = \{ d: d_i(y) = b_i y^k, ~ b_i > 0, \forall s \in S\}$.
Also, let $D_{convex}= \{d: d_i \textrm{ is convex} ~ \forall i \in S\}$.

\begin{co}
\label{poaresults}
\textbf{Correlation Induces Anarchy:}
Let  $z_e(\bx_e) =\left(\sum_{t \in T} x_{e,t}^n\right)^{\frac{1}{n}}$, $\Psi_{e,t}(\bx_e)  = \frac{ x_{e,t}^n}{ \left(\sum_{j \in T} x_{e,j}^n\right)}$,
 $y_s(\brho_s)= \left( \sum_{t \in T} R_{s,t}^m
 \right)^{\frac{1}{m}}$,
and  $\Phi_{s,t}(\brho_s)= \frac{1}{N_T}$, then we have
\begin{enumerate}
\item
$\rho(\mathcal{G}_{all}, \mathcal{C}_{mon}, \mathcal{D}_{convex}, \Psi, \Phi, \mathcal{M}_{ind}) = 1.$
\item
$\rho(\mathcal{G}_{all}, \mathcal{C}_{N_T}, \mathcal{D}_{convex}, \Psi, \Phi, \mathcal{M}_{c}) = 1.$
\item
$\rho(\mathcal{G}_{all}, \mathcal{C}_{mon}, \mathcal{D}_{convex},
\Psi, \Phi, \mathcal{M}_{c}) > 1$ for large values of $m$ and $n$. \newline
In fact,
$\rho(\mathcal{G}_{all}, \mathcal{C}_{1}, \mathcal{D}_{2}, \Psi, \Phi, \mathcal{M}_{c}) > \frac{1+N_T}{5}$.
\item
$\rho(\mathcal{G}_{dsw}, \mathcal{C}_{mon}, \mathcal{D}_{convex},
\Psi, \Phi, \mathcal{M}_{c}) > 1$ for large values of $m$ and $n$.
\end{enumerate}
\end{co}
\proof Let $ c \in \mathcal{C}_{mon}$ i.e. $c_e(x)=a_e x^k$ for $a_e >0$ for all $e \in E$, therefore,
$\int \frac{c_e(x)}{x} dx = \int a_e x^{k-1} ~dx = a_e \frac{1}{k} x^k = \frac{1}{k} c_e(x)$.
Also, $d \in \mathcal{D}_{convex}$.
Now, since the sources are independent (i.e. $\mathcal{R_{SW}} \in \mathcal{M}_{ind}$),
from Theorem \ref{waldropopteq} and Corollary \ref{optnc} it follows
that a Wardrop flow-rate for instance
 $(G, c, d, \mathcal{R_{SW}}, \Psi, \Phi)$ is also an OPT flow-rate for the instance
 $(G, c, d, \mathcal{R_{SW}})$ which implies that
$\rho(\mathcal{G}_{all}, \mathcal{C}_{mon}, \mathcal{D}_{convex},
\Psi, \Phi, \mathcal{M}_{ind}) = 1.$ \\

Even if the sources are correlated, when we have $k =N_T$, we have $ N_T \int \frac{c_e(x)}{x} dx = c_e(x)$
and using Theorem \ref{waldropopteq},  a Wardrop flow-rate for instance
 $(G, c, d, \mathcal{R_{SW}}, \Psi, \Phi)$ is also an OPT flow-rate for the instance
 $(G, c, d, \mathcal{R_{SW}})$ which implies that
\begin{equation*}
\rho(\mathcal{G}_{all}, \mathcal{C}_{N_T}, \mathcal{D}_{convex}, \Psi, \Phi, \mathcal{M}_{c}) = 1.
\end{equation*}
We prove $\rho(\mathcal{G}_{all}, \mathcal{C}_{1}, \mathcal{D}_{2}, \Psi, \Phi, \mathcal{M}_{c}) > \frac{1+N_T}{5}$
and consequently
\begin{equation*}
\rho(\mathcal{G}_{all}, \mathcal{C}_{mon}, \mathcal{D}_{convex}, \Psi, \Phi, \mathcal{M}_{c}) > 1,
\end{equation*}
 by explicitly
constructing an example as provided in Figure \ref{poafig01}.
All sources are identical with entropy $h$, therefore, $\mathcal{R_{SW}} \in \mathcal{M}_c$.
Let $d_s(y)=C_1 y^2$ for all $s \in S$, therefore, $d \in \mathcal{D}_{2}$, and the
edge cost functions,
$c_e(x)=x$ except for the edge $(u,v)$ for which
$c_e(x)=C_2~x$. Therefore, $ c \in \mathcal{C}_{1}$.
Let us consider the following flow-rate $(f,R)$
\begin{eqnarray*}
R_{1,t} &=& h ~~ \forall t \in T \\
R_{s,t}&=& 0 ~~ \forall s \in S-\{1\}, t \in T \\
f_{(1,t)} &=& h ~~  \forall t \in T \text{~over dotted edges in Figure \ref{poafig01}}\\
f_{P} &=&0 ~~~ \forall P \in \mathcal{P}_t-\{ (1,t)\} , t \in T.
\end{eqnarray*}
Clearly, $(f,R)$ is feasible for the instance $(G,c,d,\mathcal{R_{SW}})$.
We claim that
$(f,R)$ is a Wardrop flow-rate for the instance $(G,c,d,\mathcal{R_{SW}}, \Psi, \Phi)$ when $\frac{2C_1h}{N_T} \leq 1+C_2$.
To see this, first note that $(f,R)$ satisfies the Conditions (1) and (2) in the definition
of Wardrop flow-rate (Definition \ref{waldropdef}) for the instance $(G,c,d,\mathcal{R_{SW}}, \Psi, \Phi)$.
We will now check the conditions (3) and (4) in Definition \ref{waldropdef}.
Note that $\Psi_{e,t}(\bx_e) = \frac{1}{N_T}$
whenever $x_{e,t}=x$ for all $ t \in T$ for some $ x > 0$ and by continuity this is true even if $x = 0$. Therefore,
\begin{align*}
C_{(1,t)}(f) &= \sum_{e \in \{(1,t)\}}  \frac{c_e(z_e)
  \Psi_{e,t}(\bx_e)}{x_{e,t}}  = \frac{h ~. ~ 1}{h} = 1,\\
C_{(1,u,v,t)}(f) &= \sum_{e \in \{(1,u), (u,v), (v,t)\}}  \frac{c_e(z_e) \Psi_{e,t}(\bx_e)}{x_{e,t}} \\
&= \lim_{x \longrightarrow 0} \bigg{[}\frac{x ~. ~ (1/N_T)}{x} +
 \frac{C_2 x ~. ~(1/N_T)}{x} +  \frac{x ~. ~ 1}{x}\bigg{]}\\
& = 1 + \frac{1 + C_2}{N_T}, \text{~and similarly}\\
C_{(s,u,v,t)}(f) & =  1 + \frac{1 + C_2}{N_T},~ s \in S-\{1\}.
\end{align*}
Clearly, the condition (3) is satisfied as $C_{(1,t)}(f) <
C_{(1,u,v,t)}(f)$.
Also,
\begin{eqnarray*}
\frac{\partial C_S^{(t)}(R)}{\partial R_{i,t}}  &=& \frac{1}{N_T} d_{i}^{'}(y_i) y_{i,t}^{'}(\brho_i) \\
 &=& \frac{1}{N_T} 2 C_1 y_i  y_{i,t}^{'}(\brho_i) \\
&=& \frac{2C_1}{N_T} y_i^2 \frac{R_{i,t}^{m-1}}{\sum_{j \in T} R_{i,j}^m}\\
&=&  \frac{2C_1}{N_T} \left(\sum_{j \in T} R_{i,j}^m\right)^{2/m}
\frac{R_{i,t}^{m-1}}{\sum_{j \in T} R_{i,j}^m}.\\
\therefore ~ \frac{\partial C_S^{(t)}(R)}{\partial R_{1,t}} &=&
\frac{2C_1}{N_T} \left(N_T h^m\right)^{2/m} \frac{h^{m-1}}{N_T h^m} \\
& =& \frac{2C_1h}{N_T^2} ~~~ \textrm{ as $m \longrightarrow \infty$ }
~~ \textrm{  and  } \\
\frac{\partial C_S^{(t)}(R)}{\partial R_{s,t}} &\geq& 0 , \forall s
\in S-\{1\}.
\end{eqnarray*}

Therefore, when  $\frac{2C_1h}{N_T} \leq 1+C_2$, we get
\begin{eqnarray*}
C_{(1,t)}(f)+\frac{\partial C_S^{(t)}(R)}{\partial R_{1,t}} &\leq&
C_{(s,u,v,t)}(f) + \frac{\partial C_S^{(t)}(R)}{\partial R_{s,t}} \\
& & ~~~ \forall s \in S-\{1\}
\end{eqnarray*}
which implies that the condition (4) is also
satisfied. Thus, $(f,R)$ is indeed a Wardrop flow-rate for the
instance $(G,c,d,\mathcal{R_{SW}}, \Psi, \Phi)$.
Further,
\begin{eqnarray*}
C(f,R)& =& \sum_{e \in \cup_{t \in T} \{ (1,t)\}} c_e(z_e)
 + \sum_{e
  \in \cup_{s \in S} \{ (s,u)\}} c_e(z_e) \\
& & ~~  + c_{(u,v)}(z_{(u,v)}) + \sum_{e \in \cup_{t \in T} \{ (v,t)\}}
c_e(z_e) + \sum_{s \in S} d_s(y_s)\\
& = & N_T h + 0 + 0 + 0 + C_1 \left(N_T h^m\right)^{2/m} \\
&=& N_T h+ C_1h^2 \textrm{   as $m \longrightarrow \infty$}.\\
\end{eqnarray*}
Now let us consider another flow-rate $(f^{*},R^{*})$
\begin{align*}
R^{*}_{s,t}&= \frac{h}{N_S}~~ \forall s \in S, t \in T \\
f^{*}_{(1,t)} &= 0 ~  \forall t \in T,\textrm{and~} \\
f^{*}_{(s,u,v,t)} &  = \frac{h}{N_S} ~~~ \forall s \in S , t \in T.
\end{align*}
Clearly, $(f^{*},R^{*})$ is feasible for
the instance $(G,c,d,\mathcal{R_{SW}})$.
Further,
\begin{eqnarray*}
C(f^{*},R^{*}) &=& \sum_{e \in \cup_{t \in T} \{ (1,t)\}}
c_e(z_e^{*}) 
 + \sum_{e \in \cup_{s \in S} \{ (s,u)\}} c_e(z_e^{*}) +
c_{(u,v)}(z_{(u,v)}^{*}) \\
& & + \sum_{e \in \cup_{t \in T} \{ (v,t)\}}
c_e(z_e^{*}) + \sum_{s \in S} d_s(y_s^{*})\\
&=& 0 + N_S \left(N_T (\frac{h}{N_S})^n\right)^{1/n} + C_2 (N_T
h^n)^{1/n} \\
& & +
N_T h  + N_S C_1 \left(N_T (\frac{h}{N_S})^m\right)^{2/m} \\
&=& h (1+C_2+N_T) + \frac{C_1h^2}{N_S} \\
& & ~~ \textrm{   as $m
  \longrightarrow \infty, n \longrightarrow \infty$}.
\end{eqnarray*}

Thus, when $\frac{1+C_2}{C_1} < h~(1-\frac{1}{N_S})$, we have
$C(f^{*},R^{*}) < C(f,R)$. As
$OPT(G,c,d,\mathcal{R_{SW}}) \leq
C(f^{*},R^{*})$,
this implies that the POA is greater than one. \newline
In particular,
$$ \rho(\mathcal{G}_{all}, \mathcal{C}_{1}, \mathcal{D}_{2}, \Psi,
\Phi, \mathcal{M}_{c}) > \frac{C_1+\frac{N_T}{h}}{\frac{1+C_2+N_T}{h}
  + \frac{C_1}{N_S}}.$$
Now, take $h=1, N_S=N_T > 4, 1+C_2=3N_T,C_1=N_T^2$,
and note that
\begin{equation*}
\frac{2C_1h}{N_T}= 2N_T < 3N_T = 1+C_2,
\end{equation*}
as well as,
\begin{equation*}
\frac{1+C_2}{C_1} = \frac{3}{N_T} < (1-\frac{1}{N_T}) =
(1-\frac{1}{N_S}) ~~~ \textrm{as} ~~ N_T > 4.
\end{equation*}
Therefore,
we get
\begin{equation*}
\rho(\mathcal{G}_{all}, \mathcal{C}_{1}, \mathcal{D}_{2},
\Psi, \Phi, \mathcal{M}_{c}) >\frac{1+N_T}{5}.
\end{equation*}
This is near tight as will be evident from Theorem \ref{poa-ubound}.

To establish (4), we will prove a stronger result, $\rho(\mathcal{G}_{dsw},
\mathcal{C}_{3}, \mathcal{D}_{3}, \Psi, \Phi, \mathcal{M}_{c})
> 1$, by constructing an example as described below.
As shown in Figure \ref{poafig02}, there are two sources and two terminals which are directly connected
to each source. Both sources are identical with entropy $1$,
$d_1(y)=C_1y^3, d_2(y)=C_2y^3$ with $C_1, C_2 > 0, C_1 \neq C_2$ and
$c_e(x)=x^3$ for all edges. We now outline the argument that shows that the POA $>$ 1.

First, observe that the instance is symmetric with respect to
terminals and all cost functions are strictly convex. Therefore the OPT flow rate for the instance, denoted $(f^{*},R^{*})$ is such that $R_{s,t_1}^{*}=R_{s,t_2}^{*}$ for $s=1,2$.
Next, by the characterization as per Theorem
\ref{waldropopteq}, the Wardrop flow-rate, denoted $(f,R)$ is an OPT
flow-rate for $\tilde{c}_e(x) = \frac{2}{3}x^3$ with the source cost
functions remaining the same. This new instance with $\tilde{c}_e(x) =
\frac{2}{3}x^3$ is also symmetric with respect to the terminals and
the cost functions
remain strictly convex. Therefore we conclude that for the Wardrop flow-rate as well $R_{s,t_1}=R_{s,t_2}$ for $s=1,2$.
Let $R_{1,t_1}=R_{1,t_2}=h$ and $R_{1,t_1}^{*}=R_{1,t_2}^{*}=h^{*}$.
Using the properties of Wardrop flow-rate and OPT flow rate as per
Condition (2) in Theorem \ref{fourcondopt},  we have
$ R_{2,t_1}=R_{2,t_2}=1-h$ and $R_{2,t_1}^{*}=R_{1,t_2}^{*}=1-h^{*}$.
We argue below that $ h \neq h^{*}$. Consequently,
by uniqueness of the OPT flow-rate (due to strict convexity of the objective function) we will have $C(f,R) > C(f^{*},R^{*})$ implying
$\rho(\mathcal{G}_{dsw}, \mathcal{C}_{3}, \mathcal{D}_{3}, \Psi, \Phi,
\mathcal{M}_{c}) > 1$. We have, for $t=t_1,t_2$,
\begin{align*}
\frac{\partial C_S^{(t)}(R)}{\partial R_{1,t}} &= \frac{1}{N_T}
d_{1}^{'}(y_1) y_{1,t}^{'}(\brho_1)\\
&= \frac{3}{2} C_1 y_1^2 y_1
\frac{R_{1,t}^{m-1}}{\sum_{j=1}^2 R_{1,j}^m} \\
& = \frac{3}{4} C_1 h^2 \text{~ as $m \rightarrow \infty$}.
\end{align*}

Similarly,
\begin{equation*}
\frac{\partial C_S^{(t)}(R)}{\partial R_{2,t}}
=\frac{3}{4} C_2 (1-h)^2.
\end{equation*}
By the definition of Wardrop flow-rate, we have
\begin{equation*}
f_{(1,t)}= h, ~~~~ f_{(2,t)}=(1-h).
\end{equation*}
 Thus,
\begin{equation*}
C_{(1,t)}(f)= h^2, ~~~~ C_{(2,t)}(f)=(1-h)^2.
\end{equation*}
Further,
\begin{equation*}
\frac{\partial C_S^{(t)}(R)}{\partial R_{1,t}} + C_{(1,t)}(f) =
\frac{\partial C_S^{(t)}(R)}{\partial R_{2,t}} + C_{(2,t)}(f)
\end{equation*}
implies that
\begin{equation*}
 \frac{3}{4} C_1 h^2 + h^2 =\frac{3}{4} C_2 (1-h)^2 + (1-h)^2.
\end{equation*}
Therefore,
\begin{equation*}
\frac{h}{1-h} = \sqrt{\frac{\frac{3}{4} C_2+1}{\frac{3}{4}C_1+1}}.
\end{equation*}
Now, from Theorem \ref{waldropopteq}, $(f^{*}, R^{*})$ is a Wardrop
flow-rate
for the instance where everything remains the same except for the edge
cost functions which are now $\frac{3}{2} x^3$ instead of $x^3$ and
performing the similar calculations as above for $(f,R)$, we obtain
\begin{equation*}
 \frac{h^{*}}{1-h^{*}} = \sqrt{\frac{\frac{3}{4} C_2+\frac{3}{2}}{\frac{3}{4}
    C_1+\frac{3}{2}}}.
\end{equation*}
 Clearly, since $C_1 \neq C_2$, we get $ h \neq
h^{*}$. In particular, take $C_1=4, C_2=8$, then $h=0.5695$ and
$h^{*}=0.5635$.
Thus, $C(f,R)=1.9061, C(f^{*},R^{*})=1.9052$ implying that $POA \geq
1.004 > 1$, in this example. \qed
\endproof

Note that while constructing the above examples
the source cost splitting function we have used is $\Phi_{s,t}(\brho_s)=1/N_T$.
Further, for the same mechanism, Corollary \ref{poaresults}(2) provides an example of edge cost functions
that gives a POA of one, and possibly this is the only choice giving POA one.
Before considering another reasonable splitting mechanism, we first establish an
upper bound which is nearly attainable by instance given in Figure \ref{poafig01}.
\begin{theo}
\label{poa-ubound}
Let  $z_e(\bx_e) =\left(\sum_{t \in T} x_{e,t}^n\right)^{\frac{1}{n}}, \Psi_{e,t}(\bx_e)  = \frac{ x_{e,t}^n}{ \left(\sum_{j \in T} x_{e,j}^n\right)}$ and $  \Phi_{s,t}(\brho_s)= \frac{1}{N_T}$. Then,
\begin{equation*}
\rho(\mathcal{G}_{all}, \mathcal{C}_{k}, \mathcal{D}_{convex}, \Psi, \Phi, \mathcal{M}_{c}) \leq \max\{ \frac{N_T}{k}, \frac{k}{N_T}\}.
\end{equation*}
\end{theo}

\proof
As in the proof of Theorem \ref{waldropopteq},
we have,
$ C_P(f)
 = \frac{1}{N_T} \sum_{e \in P}  \tilde{c}_e^{'}(z_e)
 z_{e,t}^{'}(x_e)$ and
$ C_{P_i}(f) + \frac{\partial C_S^{(t)}(R)}{\partial R_{i,t}} = \frac{1}{N_T} \left[ \sum_{e \in P_i}  \tilde{c}_e^{'}(z_e) z_{e,t}^{'}(\bx_e)+
d_{i}^{'}(y_i) y_{i,t}^{'}(\brho_i)\right].$ \\
Let  $(f,R)$ be a Wardrop flow-rate  and $(f^{*},R^{*})$ be OPT for $(G,c,d,\mathcal{R_{SW}})$
respectively.
Further,
let
$\tilde{c}_e(x) = N_T \int \frac{c_e(x)}{ x} dx = N_T \int a_e x^{k-1} dx = \frac{N_T}{k} a_e x^k$.
Now,
\begin{align*}
C(f,R) & = \sum_{e \in E} c_e(z_e) + \sum_{ s \in S} d_s(y_s) = \sum_{e \in E} a_e z_e^{k} + \sum_{ s \in S} d_s(y_s)
\end{align*}
and
\begin{align*}
C(f^{*},R^{*})&  = \sum_{e \in E} c_e(z_e^{*}) + \sum_{ s \in S} d_s(y_s^{*})\\
& = \sum_{e \in E} a_e (z_e^{*})^{k} + \sum_{ s \in S} d_s(y_s^{*})
\end{align*}
Let us first consider the case where $N_T \geq k$ i.e. $ 1 \leq
\frac{N_T}{k}$.
\begin{eqnarray*}
C(f,R)  &=& \sum_{e \in E} a_e z_e^{k} + \sum_{ s \in S} d_s(y_s)  \\
& \leq& \sum_{e \in E} \frac{N_T}{k} a_e z_e^{k} + \sum_{ s \in S} d_s(y_s) \\
&=&  \sum_{e \in E} \tilde{c}_e(z_e) + \sum_{ s \in S} d_s(y_s).
\end{eqnarray*}
Now,  from Theorem \ref{waldropopteq}, $(f,R)$ is OPT for $(G,\tilde{c},d,\mathcal{R_{SW}})$ and because
$(f^{*},R^{*})$ is feasible for  $(G,\tilde{c},d,\mathcal{R_{SW}})$
we get
\begin{eqnarray*}
\sum_{e \in E} \tilde{c}_e(z_e) + \sum_{ s \in S} d_s(y_s)
& \leq & \sum_{e \in E}  \tilde{c}_e(z_e^{*}) +  \sum_{ s \in S} d_s(y_s^{*}) \\
&=& \sum_{e \in E} \frac{N_T}{k} a_e (z_e^{*})^{k} + \sum_{ s \in S} d_s(y_s^{*}) \\
&\leq&  \frac{N_T}{k}  \left[ \sum_{e \in E} a_e (z_e^{*})^{k} +  \sum_{ s \in S} d_s(y_s^{*}) \right] \\
&= &\frac{N_T}{k} ~ C(f^{*},R^{*}).
\end{eqnarray*}
Therefore,
\begin{equation*}
 \frac{ C(f,R) }{C(f^{*},R^{*})} \leq \frac{N_T}{k}.
\end{equation*}

Similarly, for the case when  $N_T \leq k$ i.e. $ 1 \geq
\frac{N_T}{k}$,
\begin{eqnarray*}
C(f,R) & =& \sum_{e \in E} a_e z_e^{k} + \sum_{ s \in S} d_s(y_s) \\
& =& \frac{k}{N_T} \left[ \sum_{e \in E} \frac{N_T}{k} a_e z_e^{k} +  \sum_{ s \in S} \frac{N_T}{k} d_s(y_s) \right]\\
& \leq& \frac{k}{N_T} \left[ \sum_{e \in E} \frac{N_T}{k} a_e z_e^{k} + \sum_{ s \in S} d_s(y_s)\right]\\
& =& \frac{k}{N_T} \left[ \sum_{e \in E} \tilde{c}_e(z_e) + \sum_{ s \in S} d_s(y_s) \right] \\
\end{eqnarray*}

Now,  from Theorem \ref{waldropopteq}, $(f,R)$ is OPT for $(G,\tilde{c},d,\mathcal{R_{SW}})$ and because
$(f^{*},R^{*})$ is feasible for  $(G,\tilde{c},d,\mathcal{R_{SW}})$
we get
\begin{eqnarray*}
 \sum_{e \in E} \tilde{c}_e(z_e) + \sum_{ s \in S} d_s(y_s)
&\leq&  \sum_{e \in E}  \tilde{c}_e(z_e^{*}) + \sum_{ s \in S} d_s(y_s^{*})  \\
& =& \sum_{e \in E} \frac{N_T}{k} a_e (z_e^{*})^{k} + \sum_{ s \in S} d_s(y_s^{*}) \\
&\leq& \sum_{e \in E} a_e (z_e^{*})^{k} + \sum_{ s \in S} d_s(y_s^{*}) \\
&=& C(f^{*},R^{*})
\end{eqnarray*}
Therefore,
\begin{align*}
\frac{ C(f,R) }{C(f^{*},R^{*})} \leq \frac{k}{N_T}.
\end{align*}
\endproof

Now we consider another splitting mechanism $\Phi$ 
that looks more like the edge cost splitting mechanism $\Psi$.
Specifically, take $y_s(\brho_s) = \left( \sum_{t \in T} (R_{s,t})^m \right)^{\frac{1}{m}}$ and
$\Phi_{i,t}(\brho_i) = \frac{(R_{i,t})^m}{\sum_{j \in T}
  (R_{i,j})^m}$.
Let us first note the generalization of Corollary \ref{poaresults}(1)
for any source cost splitting mechanism $\Phi$. Proof is esentially the
same as before. The condition (2) in the definition of Wardrop
flow-rate as well as OPT flow-rate renders all the rates to be equal to their corresponding
entropies and consequently the condition (4) need not be
checked.
\begin{lemma}
Let  $z_e(\bx_e) =\left(\sum_{t \in T} x_{e,t}^n\right)^{\frac{1}{n}}$, $\Psi_{e,t}(\bx_e)  = \frac{ x_{e,t}^n}{ \left(\sum_{j \in T} x_{e,j}^n\right)}$,
 and  $\Phi_{s,t}(\brho_s)$ be any source cost splitting function, then we have
\begin{equation*}
\rho(\mathcal{G}_{all}, \mathcal{C}_{mon}, \mathcal{D}_{convex}, \Psi, \Phi, \mathcal{M}_{ind}) = 1.
\end{equation*}
\end{lemma}
Now, we will argue that with $y_s(\brho_s) = \left( \sum_{t \in T} (R_{s,t})^m \right)^{\frac{1}{m}}$ and
$\Phi_{i,t}(\brho_i) = \frac{(R_{i,t})^m}{\sum_{j \in T}
  (R_{i,j})^m}$ we have \newline 
$\rho(\mathcal{G}_{dsw}, \mathcal{C}_{mon}, \mathcal{D}_{convex},
\Psi, \Phi, \mathcal{M}_{c}) > 1$ 
for large values of $m$ and
$n$. Let us consider the same example as in Figure \ref{poafig02}
but with the new source cost splitting mechanism.
First, note that OPT flow-rate is independent of the choice of cost splitting functions and
the previously calculated OPT flow-rate for this instance
$(f^{*},R^{*})$ is given by 
\begin{eqnarray*}
R_{1,t}^{*}=f_{(1,t)}^{*}=h^{*} , ~~~~ \textrm{and} \\
R_{2,t}^{*}=f_{(2,t)}^{*}=1-h^{*}.
\end{eqnarray*}
We will argue that this is not a
Wardrop flow-rate and since the OPT flow-rate
is unique (by strict convexity) we will obtain $POA >1$.
After some simple calculations we get
\begin{align*}
 \frac{\partial C_S^{(t)}(R)}{\partial R_{i,t}} 
= d_i^{'}(y_i)
\frac{y_i}{R_{i,t}} \Phi_{i,t}^2(\brho_i)
+ m \frac{d_i(y_i)}{R_{i,t}}
\Phi_{i,t}(\brho_i) \left(1-\Phi_{i,t}(\brho_i)\right).
\end{align*}
Therefore,
\begin{eqnarray*}
\frac{\partial C_S^{(t)}(R^{*})}{\partial R_{1,t}}= (m+3) (N_T)^{3/m}
\frac{C_1}{4} (h^{*})^2 ~~~~ \textrm{and} \\
\frac{\partial C_S^{(t)}(R^{*})}{\partial R_{2,t}}= (m+3) (N_T)^{3/m}
\frac{C_2}{4} (1-h^{*})^2.
\end{eqnarray*}
Also, $C_{(1,t)}(f^{*}) = (h^{*})^2$ and $C_{(2,t)}(f^{*}) =
(1-h^{*})^2$. Note that $N_T=2$ in this example.
Now, with $C_1=4, C_2=8$, we have $h^{*}=0.5635$ and therefore
\begin{eqnarray*}
 \frac{C_{(1,t)}(f^{*})+\frac{\partial C_S^{(t)}(R^{*})}{\partial
    R_{1,t}}}{C_{(2,t)}(f^{*})+\frac{\partial
    C_S^{(t)}(R^{*})}{\partial R_{2,t}}} 
&=&\frac{(h^{*})^2 + (m+3) (N_T)^{3/m}
\frac{C_1}{4} (h^{*})^2}{(1-h^{*})^2 + (m+3) (N_T)^{3/m}
\frac{C_2}{4} (1-h^{*})^2} \\
&=& \frac{(m+3) (N_T)^{3/m} + 1}{2(m+3) (N_T)^{3/m} + 1} ~
\frac{0.5635^2}{(1-0.5635)^2} \\
&=& \frac{1}{2}
\frac{0.5635^2}{(1-0.5635)^2}\\
& =& 0.8333 \neq 1 \textrm{~as $m \rightarrow \infty$}.
\end{eqnarray*}

\begin{theo}
Let  $z_e(\bx_e) =\left(\sum_{t \in T} x_{e,t}^n\right)^{\frac{1}{n}}$, $y_s(\brho_s) = \left( \sum_{t \in T} (R_{s,t})^m \right)^{\frac{1}{m}}$, $\Psi_{e,t}(\bx_e)  = \frac{ x_{e,t}^n}{ \left(\sum_{j \in T} x_{e,j}^n\right)}$, and
$\Phi_{i,t}(\brho_i) = \frac{(R_{i,t})^m}{\sum_{j \in T}
  (R_{i,j})^m}$ for large values of $m$ and $n$, then we have
\begin{equation*}
\rho(\mathcal{G}_{dsw}, \mathcal{C}_{mon}, \mathcal{D}_{convex},\Psi, \Phi, \mathcal{M}_{c}) > 1.
\end{equation*}
\end{theo}

\section{ Future Directions}
 \label{concl-sec}
In this work, we have initiated a study of the inefficiency brought forth by
the lack of regulation in the multicast of \emph{multiple correlated sources}.
We have established the foundations of
the framework by providing the first set
of technical results that characterize the equilibrium among terminals, when
they act selfishly trying to minimize their individual costs without any regard to
social welfare, and its relation to the socially optimal solution.
Our work leaves out several important open problems that deserve theoretical
investigation and analysis. We discuss some of these interesting problems in the following.

\paragraph*{Network Information Flow Games: From Slepian-Wolf  to Polymatroids:}
It is interesting to note that all the results presented
in this chapter naturally extends to a large class of network information flow problems
where the entropy is replaced by any rank function (ref. Chapter 10 in \cite{GLS93}) and
equivalently conditional entropy is replaced by any
supermodular function. This is because
the only special property of conditional entropy used in our analysis is its
supermodularity. Polytopes described by such rank functions
are called \textit{contra-polymatroids} and the SW polytope is an example.
Therefore, by abstracting the
network coding scenario to this more general setting,
we can obtain a nice class of multi-player games with compact representations,
which we call \textit{Network Information Flow Games}.
It would be interesting to study these games further and investigate the emergence of
 practical and meaningful scenarios beyond network coding.
Furthermore, the network coding scenario where the terminals do
not necessarily want to reconstruct all the sources should also be interesting
to analyze.

\paragraph*{Dynamics of Wardrop Flow-Rate:}
Can we design a noncooperative decentralized algorithm that steers flows and rates in way that
converges to a Wardrop flow-rate? What about such an algorithm which runs in polynomial time?
A first approach could be to consider an algorithm where each terminal greedily allocates
 rates and flows by calculating marginal costs at each step. The
 following theorem, which follows from an approach similar to that in the
 proof of Theorem \ref{fourcondopt}, provides some intuition on why such a greedy
approach might work, as per the relationship between Wardrop and OPT according to Theorem \ref{waldropopteq}.
\begin{theo}
\label{opt-greedy-rate}
Let $(f,R)$  be an  OPT flow-rate for instance $(G,c,d, \mathcal{R_{SW}})$ and
define $h_{s,t} := d_s^{'}(y_s)y_{s,t}^{'}(\brho_s) + \lambda_{s,t}$ for $s \in S, t \in T$, where
$\lambda_{s,t}$'s are dual variables satisfying KKT conditions \ref{kkt-flow}, \ref{kkt-flow-rate}.
Further, let $\sigma : T \times S \longrightarrow S$ be defined such that
$0 < h_{\sigma(t,1),t} < h_{\sigma(t,2),t} < \dots < h_{\sigma(t,N_s),t}$. Then,
\begin{eqnarray*}
\sum_{i=1}^k R_{\sigma(t,i),t} = H(X_{\sigma(t,1)},X_{\sigma(t,2)}, \dots, X_{\sigma(t,k)}) 
 ~~ \textrm{for $k = 1, \dots, N_s$.}
\end{eqnarray*}
\end{theo}

\paragraph*{ Better bounds on POA:}
Although we have provided explicit examples where correlation brings more anarchy, as well as,
an upper bound on POA which is nearly achievable, we believe that more detailed analysis is necessary.
An important approach in this direction would be to characterize exactly how the POA depends
on structure of SW region i.e. to analyze the finer details on how correlation among sources
changes POA, even in the case of two sources.
Further, other interesting splitting mechanisms should also be
studied.

\paragraph*{Capacity Constraints and Approximate Wardrop Flow-Rates:} One immediate direction of investigation
could be to consider the scenario where there is a capacity constraint on each edge i.e. the
maximum amount of flow that can be sent through that edge. Another interesting problem is
to investigate the sensitivity of the implicit assumption in our analysis that terminals
can evaluate various quantities, and in particular the marginal costs, with arbitrary
precision. This can be achieved by formulating a notion of approximate Wardrop flow-rate, where terminals can
distinguish quantities only when they differ significantly.


\bibliographystyle{abbrv}
\bibliography{refthesis,tip}

\end{document}